\documentclass[journal]{IEEEtran}
\IEEEoverridecommandlockouts
\usepackage{booktabs}
\usepackage{threeparttable}
\usepackage{amsmath,graphicx,cite,amssymb}
\usepackage{amsfonts,multirow,bm,array,setspace}
\usepackage{graphicx,float,cite,amssymb}
\usepackage{multirow,bm,array,setspace}
\usepackage{textcomp}
\usepackage{psfrag}
\usepackage{color}
\usepackage{algorithm}
\usepackage{algorithmic}
\usepackage{bbm}
\usepackage{cuted}
\usepackage{subfigure}
\usepackage{flafter}
\usepackage{makecell}
\usepackage{xcolor}
\usepackage{xpatch}

\newtheorem{proposition}{Proposition}

\newtheorem{lemma}{Lemma}
\newtheorem{theorem}{Theorem}

\newtheorem{remark}{Remark}

\def\BibTeX{{\rm B\kern-.05em{\sc i\kern-.025em b}\kern-.08em
    T\kern-.1667em\lower.7ex\hbox{E}\kern-.125emX}}

\makeatletter
\newcommand{\distas}[1]{\mathbin{\overset{#1}{\kern\z@\sim}}}%
\newsavebox{\mybox}\newsavebox{\mysim}
\newcommand{\distras}[1]{%
	\savebox{\mybox}{\hbox{\kern1pt$\scriptstyle#1$\kern1pt}}%
	\savebox{\mysim}{\hbox{$\sim$}}%
	\mathbin{\overset{#1}{\kern\z@\resizebox{\wd\mybox}{\ht\mysim}{$\sim$}}}%
}

\ExplSyntaxOn
\cs_new:Npn \bibColoredItems #1#2
  {
    \clist_map_inline:nn {#2} { \cs_new:cpn {bib@colored@##1} {#1} } 
  }
\ExplSyntaxOff

\newcommand\bib@setcolor[1]{%
  \ifcsname bib@colored@#1\endcsname
    \expandafter\color\expandafter{\csname bib@colored@#1\endcsname}
  \else
    \normalcolor
  \fi
}

\xpatchcmd\@bibitem
  {\item}
  {\bib@setcolor{#1}\item}
  {}{\PatchFailed}

\xpatchcmd\@lbibitem
  {\item}
  {\bib@setcolor{#2}\item}
  {}{\PatchFailed}
  
\begin{document}

\title{Fractional Programming for Stochastic Precoding over Generalized Fading Channels}

\author{
    \IEEEauthorblockN{
    Wenyu Wang, \IEEEmembership{Graduate Student Member,~IEEE} and Kaiming Shen, \IEEEmembership{Senior Member,~IEEE}
   } 
\thanks{Accepted to IEEE Transactions on Signal Processing on 10 February 2026. The work was supported in part
by the NSFC under Grant 12426306 and in part by Guangdong Basic and Applied Basic Research under Grant 2023B0303000001. \emph{(Corresponding author: Kaiming Shen.)}

The authors are with School of Science and Engineering, The Chinese University of Hong Kong (Shenzhen), China (e-mail: wenyuwang@link.cuhk.edu.cn; shenkaiming@cuhk.edu.cn). 
    }
}

\maketitle

\begin{abstract}
This paper seeks an efficient algorithm for stochastic precoding to maximize the long-term average weighted sum rates throughout a multiple-input multiple-output (MIMO) network. Unlike many existing works that assume a particular probability distribution model for fading channels (which is typically Gaussian), our approach merely relies on the first and second moments of fading channels. For the stochastic precoding problem, a naive idea is to directly apply the fractional programming (FP) method to the data rate inside the expectation; it does not work well because the auxiliary variables introduced by FP are then difficult to decide. To address the above issue, we propose using a lower bound to approximate the expectation of data rate. This lower bound stems from a nontrivial use of the matrix FP, and outperforms the existing lower bounds in that it accounts for generalized fading channels whose first and second moments are known. The resulting approximate problem can be efficiently solved in closed form in an iterative fashion. Furthermore, for large-scale MIMO, we improve the efficiency of the proposed algorithm by eliminating the large matrix inverse. Simulations show that the proposed stochastic precoding method outperforms the benchmark methods in both Gaussian and non-Gaussian fading channel cases.
\end{abstract}
\begin{IEEEkeywords}
Stochastic precoding, multiple-input multiple-output (MIMO), first and second moments, long-term average weighted sum rates, matrix fractional programming (FP).
\end{IEEEkeywords}

\section{Introduction}
\IEEEPARstart{S}{tochastic} precoding, a.k.a. robust beamforming, aims to combat the uncertainty of fading channels by maximizing the long-term average weighted sum rates throughout a multiple-input multiple-output (MIMO) network. While many existing works specialize to the Gaussian distribution model of fading channels, we do not assume any probability distribution model except that the first and second moments of fading channels are known. Our approach is based on a stochastic extension of the matrix fractional programming (FP) method \cite{fp_i,fp_learning}, with the data rate of each user treated as a function of random matrix ratio.

The first order of business in stochastic precoding is to model fading channels. A widely used model \cite{gaussianerror2,robust_rzf,robust_slnr,robust_mmse_cellfree1,robust_mmse_cellfree2,gaussianerror4,shi21,shi23,aal,gaussianerror1,swmmse_al,gaussianerror5,gaussianrobust} assumes fading channels to be Gaussian, so it encompasses the classic Rayleigh and Rician fading models as special cases; \cite{bernoulli1,bernoulli2,bernoulli3} further develops it into a more sophisticated Bernoulli-Gaussian fading model. With the probability distribution of fading channels available, the expectation of data rate can be computed as an explicit function of precoders, so the standard optimization methods are applicable. However, the aforementioned works are limited to the specific fading channel models they assume. To enhance flexibility, \cite{bounded&momenterror,momenterror1} no longer specify the probability distribution of fading channels $\mathbf H$, but only assume that the first moment $\mathbb E[\mathbf H]$ and the second moment $\mathbb E[\mathrm{vec}(\mathbf H)\mathrm{vec}(\mathbf H)^H]$ are known, 
where $\mathrm{vec}(\cdot)$ is the vectorization of a matrix, and $(\cdot)^H$ is the conjugate transpose, namely \emph{generalized fading channels}. The present paper adopts this model too. Another fading channel model \cite{boundederror1,robust_mmse,boundederror3,boundederror4} without specifying the distribution function is to assume the first moment $\mathbb E[\mathbf H]$ and a deviation bound $\delta>0$ so that $\|\mathbf H-\mathbb E[\mathbf H]\|_\mathrm{F}\le \delta$, where $\|\cdot\|_\mathrm{F}$ is the Frobenius norm. The authors of \cite{boundederror5set} generalize this deviation-bound model further by replacing the Frobenius norm
with other distortion metrics. Moreover, in the extreme, \cite{swmmse,spwmmse,samplingerror1,samplingerror2,samplingerror3,samplingerror4} propose model-free approaches to stochastic precoding; they entail a large number of fading channel samples either for online optimization or for neural network training.

A variety of optimization objectives have been considered in the literature for the stochastic precoding problem. Many previous works \cite{robust_rzf,robust_mmse_cellfree1,robust_mmse_cellfree2,aal,shi21,shi23,gaussianerror1,gaussianerror5,gaussianerror4,gaussianerror3,swmmse,samplingerror2,swmmse_al,spwmmse} seek to maximize the expected weighted sum of data rates throughout the network, and so does our work. For multiple-input single-output (MISO) secure transmission, \cite{momenterror1} considers maximizing the signal-to-noise (SNR) of a legitimate user while suppressing that of an eavesdropper; the SNR is approximated as the signal-to-leakage-and-noise ratio (SLNR) in \cite{robust_slnr} to ease optimization. All these references except \cite{momenterror1} require the prior knowledge of the probability distribution of fading channels. In contrast, this work assumes that only the first and second moments of fading channels are available, so the expectation of data rate is not even computable, which poses a major challenge to us. It is worth mentioning that the problem case of \cite{momenterror1} is restricted to a simple network model with the single-stream transmission toward a single receiver, so its moment-based precoding scheme does not work for our case.

Aside from the above expectation-related problem formulations, a line of previous works view stochastic precoding from a robust optimization perspective. They can be further divided into two groups. The first group \cite{robust_mmse,boundederror3,boundederror4,boundederror5set,boundederror1} maximizes the worst-case performance under the aforementioned bounded-deviation fading channel model. The other group \cite{bounded&momenterror,bernoulli1,bernoulli2,bernoulli3,samplingerror1,samplingerror4,gaussianrobust,gaussianerror2} maximizes the percentile performance based on the outage probability of data rate. This outage probability has been characterized for the Bernoulli-Gaussian fading model \cite{bernoulli1,bernoulli2,bernoulli3} and for the Rician fading model \cite{gaussianrobust}. Some other works \cite{samplingerror1,samplingerror4} sidestep the outage probability analysis and learn stochastic precoding directly by neural network with uncertainty injection.

This paper pursues the maximization of the long-term average weighted sum rates. Here and throughout, ``long-term'' refers to the timescale in which the first-order and second-order moments are evaluated for the fading channels. In other words, the data rate can be approximated as an ergodic process under this long-term setting. For this problem, \cite{swmmse} proposes a data-driven stochastic precoding algorithm that takes a large number of fading channel samples as training data. Each sample corresponds to a deterministic precoding subproblem---which is solvable by the weighted minimum mean-squared error (WMMSE) algorithm \cite{wmmse_bc,wmmse_ibc}; the main idea of \cite{swmmse} is to iteratively average out the deterministic precoding solutions. As shown in \cite{swmmse}, the iterations must converge to a stationary point of the original problem provided that the number of fading channel samples is sufficiently large. This online optimization method has evolved in some more recent works, e.g., \cite{samplingerror2} addresses the deterministic subproblems via neural networks, \cite{spwmmse} combines the deterministic solutions in a proximal fashion, and \cite{swmmse_al} tailors the online optimization to the Gaussian case. In contrast, many other works propose the offline optimization methods. However, even when fading channels are assumed to be Gaussian, the expectation of data rate is still difficult to tackle directly, so  
\cite{gaussianerror1} proposes a heuristic approximation of the original objective function, while  \cite{gaussianerror4,gaussianerror3} suggest using an upper bound to approximate it. Nevertheless, since this is a maximization problem, it is more desirable to find a lower bound for approximation. Toward this end, \cite{aal} constructs a lower bound on the expectation of weighted sum rates, but the bound is difficult to compute in practice. Some computable lower bounds are proposed in \cite{shi21,shi23} for the single-cell network with a single data stream to each user. We remark that the above bounds, either upper or lower, are all designed for the Gaussian fading channel model. In contrast, the lower bound proposed in this paper accounts for a generic fading channel model which need not be Gaussian.

The main contributions of this paper can be recognized in the following three respects:
\begin{itemize}
\item \emph{Stochastic Matrix FP:} We develop the quadratic transform and the Lagrangian dual transform in \cite{fp_i,fp_learning} to account for the stochastic matrix FP problems. This is motivated by the fact that the expectation of data rate amounts to $\mathbb E[\log|\mathbf I+\mathbf A\mathbf B^{-1}|]$, where $\mathbf A\mathbf B^{-1}$ is a random matrix ratio.
\item \emph{New Lower Bound:} Equipped with the stochastic matrix FP, we construct a lower bound on the expected weighted sum rates, which is used to approximate the stochastic precoding problem in a tractable form. This new bound is endowed with much greater generality than the existing bounds \cite{aal,shi21,shi23} in that it accounts for any fading channel models with known first and second moments.
\item \emph{Precoding Algorithm:} We show that the precoding variable can be efficiently optimized in closed form based on the first and second moments of fading channels. Furthermore, for the large-scale MIMO case, we improve the computational efficiency of the proposed algorithm by avoiding the large matrix inverse.
\end{itemize}

The remainder of this paper is organized as follows. Section \ref{sec:sysmod} describes the system model and the stochastic precoding problem. 
Section \ref{sec:sfp} contains the main results: it first discusses why the conventional way of using FP does not work for stochastic precoding, and then proposes a new way based on a generalized lower bound. Section \ref{sec:alg} specifies the above FP method for the stochastic precoding problem considered in this paper. Section \ref{sec:fast FP} discusses how to accelerate the algorithm when the transmitters have a large number of antennas. Section \ref{sec:simu} presents the simulation results. Lastly, the paper is concluded in Section \ref{sec:conclu}. 

\emph{Notation:} Matrices and vectors are in boldface. For a matrix $\mathbf{A}$, $(\mathbf A)_{i,j}$ is the $(i,j)$th entry, $\mathbf{A}^c$ is the complex conjugate, $\mathbf{A}^H$ is the conjugate transpose, $\lambda_{\max}(\mathbf{A})$ is the largest eigenvalue, and $\mathrm{vec}(\mathbf{A})$ is the vectorization of this matrix by stacking its columns, $\mathrm{Tr}(\mathbf{A})$ is the trace, $\mathbf{A}^{-1}$ is the inverse, $\sqrt{\mathbf{A}}$ is the symmetric square-root of matrix, $\Re\{\mathbf{A}\}$ retains the real part of each entry of the matrix, and $\mathrm{Diag}(\mathbf A)$ extracts the diagonal entries in a vector form. For a vector $\mathbf{a}$, $\mathrm{diag}(\mathbf{a})$ is a diagonal matrix formed by $\mathbf a$. Let $\mathbb{C}^{m\times n}$ be the set of $m\times n$ complex matrices, $\mathbb{S}^{n\times n}_{+}$ the set of $n\times n$ positive semidefinite matrices, $\mathbb{S}^{n\times n}_{++}$ the set of $n\times n$ positive definite matrices, and $\mathbf I$ the identity matrix. For two matrices $\mathbf{A}$ and $\mathbf{B}$, let $\mathbf{A}\odot\mathbf{B}$ be their Hadamard product. Moreover, for a set of subscripted variables, we use the unsubscripted symbol to denote their collection, e.g., $\mathbf V=\{\mathbf V_{jk}\}$.

\begin{table}[t]
\renewcommand\arraystretch{1.5}
\footnotesize
\centering
\caption{List of Main Variables}
\begin{tabular}{|c||l|}
\hline
\textbf{Symbol} & \textbf{Definition} \\ \hline
\hline
$L$ & number of cells \\ \hline
$K$ & number of users in each cell \\ \hline
$j,\ell$ & index of cell or BS \\ \hline
$k,s$ & index of user within each cell \\ \hline
$M_t$ & number of transmit antennas at each BS\\ \hline
$M_r$ & number of receive antennas at each user\\ \hline
$\mathbf{H}_{jk,\ell}$ & block fading channel from BS $\ell$ to user $(j,k)$ \\ \hline
$\mathbf{C}_{jk}$ & first moment of $\mathbf{H}_{jk,j}$ \\ \hline
$\mathbf{D}_{jk,\ell}$ & second moment of $\mathbf{H}_{jk,\ell}$ \\ \hline
$\mathbf{V}_{jk}$ & precoding matrix for user $(j,k)$ \\ \hline
$\mathcal{R}_{jk}$ & data rate of user $(j,k)$ \\ \hline
$\sigma^2$ & background noise power \\ \hline
$\omega_{jk}$ & date rate weight of user $(j,k)$  \\ \hline
$P$ & power constraint on each BS \\ \hline
$\mathbf{Y}$ & auxiliary variable for quadratic transform \\ \hline
$\mathbf{\Gamma}$ & auxiliary variable for Lagrangian dual transform\\ \hline
$\mathbf{Z}$ & \makecell[l]{auxiliary variable for nonhomogeneous quadratic transform } \\
\hline
$\mathbf U_{jk,\ell s}$ & shorthand for $\mathbb E\big[\mathbf{H}_{jk,\ell}\mathbf{V}_{\ell s}\mathbf{V}_{\ell s}^H\mathbf{H}_{jk,\ell}^H\big]$\\ \hline
$\bm \Lambda_{j,\ell s}$ & shorthand for $\mathbb{E}\big[\mathbf{H}_{\ell s,j}^H\mathbf{Y}_{\ell s}(\mathbf{I}+\mathbf{\Gamma}_{\ell s})\mathbf{Y}_{\ell s}^H\mathbf{H}_{\ell s,j}\big].$ \\ \hline
$\bm\Xi_j$ & shorthand for $\sum_{(\ell,s)}\omega_{\ell s}\mathbf{\Lambda}_{j,\ell s}$\\
\hline
$\alpha_j$ & largest eigenvalue of $\bm\Xi_j$ \\ \hline
\end{tabular}
\label{tab:notation}
\end{table}

\section{System Model and Problem Formulation}
\label{sec:sysmod}
Consider an $L$-cell downlink MIMO network, where each cell consists of one BS and $K$ user terminals. We refer to the BS of the $j$th cell as BS $j$, and the $k$th user terminal in the $j$th cell as user $(j,k)$, for $j,\ell=1,2,\ldots,L$ and $k=1,2,\ldots,K$, where $j$ is the main index for BS while $\ell$ is an ancillary index. Assume that each BS has $M_t$ transmit antennas while each user terminal has $M_r$ receive antennas. Let $\mathbf H_{jk,\ell}\in\mathbb C^{M_r\times M_t}$ be the block fading channel from BS $\ell$ to user $(j,k)$ in $j$th cell. These $\mathbf H_{jk,\ell}$'s are modeled as i.i.d. random variables across the different blocks. We assume that the distribution function of $\mathbf H_{jk,\ell}$ is unknown or cannot be written analytically; only the first and second moments of random channels are available:
\begin{align}
    \mathbf C_{jk} &= \mathbb E\big[\mathbf{H}_{jk,j}\big],\\
    \mathbf D_{jk,\ell} &= \mathbb E\big[\mathrm{vec}(\mathbf{H}_{jk,\ell})\mathrm{vec}(\mathbf{H}_{jk,\ell})^H\big].
\end{align}
How these moment statistics can be obtained has been extensively discussed in the literature \cite{momentest1,momentest2,momentest3}. Moreover, as pointed out in the more recent work \cite{sr-con}, the channels can be predicted by the modern tools of artificial intelligence (AI) and digital twins; in particular, \cite{sr-con} shows that the channel statistics (e.g., the first-order and second-order moments) are much easier to predict than the specific channel realization in real-world networks. The goal of stochastic precoding is to design a set of precoding matrices $\{\mathbf V_{jk}\in\mathbb C^{M_t\times M_r}:\forall (j,k)\}$ that is ``good'' for all the blocks on average.

For a particular block, the data rate of user $(j,k)$ can be computed as \cite{goldsmith2005wireless}
\begin{align}
\mathcal{R}_{jk}=\log\big|\mathbf{I}+\mathbf{H}_{jk,j}\mathbf{V}_{jk}\mathbf{V}_{jk}^{H}\mathbf{H}_{jk,j}^{H}\mathbf{F}_{jk}^{-1}\big|,
\end{align}
where
\begin{equation}
\label{F_jk}
\mathbf{F}_{jk}=\sum_{(\ell,s)\neq (j,k)}\mathbf{H}_{jk,\ell}\mathbf{V}_{\ell s}\mathbf{V}_{\ell s}^{H}\mathbf{H}_{jk,\ell}^{H}+\sigma^2\mathbf{I},
\end{equation}
with $\sigma^2$ denoting the background noise power. We use two distinct BS indices $j$ and $\ell$ to denote the cross-cell channels, e.g., the interference channel from the neighboring BS $\ell$ to the $k$th user in cell $j$ in \eqref{F_jk} is denoted by $\mathbf H_{jk,\ell}$. We aim to maximize the long-term weighted sum rates across many blocks. By the law of large numbers, the problem can be written as
\begin{subequations}\label{prob:orig}
    \begin{align}
        \underset{\mathbf{V}}{\text{maximize}}&\quad \sum^L_{j=1}\sum^K_{k=1}\mathbb{E}\Big[\omega_{jk}\mathcal{R}_{jk}\Big]\\
    \text{subject to}& \quad \sum^{K}_{k=1}\text{Tr}(\mathbf{V}_{jk}\mathbf{V}_{jk}^{H})\leq P,\  \forall j,
    \end{align}
\end{subequations}
where $\omega_{jk}\ge0$ is the rate weight assigned to user $(j,k)$ in accordance with its priority, and $P>0$ is the power constraint on each BS. The expectation in the objective function of \eqref{prob:orig} is taken over the fading channels $\mathbf H_{jk,\ell}$. For ease of reference, the main variables are listed in Table \ref{tab:notation}.

We provide two remarks about the stochastic precoding problem in \eqref{prob:orig}. First, the problem is intractable because its objective function is unknown---the expectation cannot be computed without knowing the probability distribution of $\mathbf H_{jk,\ell}$. Second, the problem is also valid in the other stochastic precoding scenarios. For instance, we could have alternatively assumed that the randomness in $\mathcal R_{jk}$ is caused by the channel estimation error; In that case, $\mathbf{C}_{jk}$ is interpreted as the estimate of $\mathbf{H}_{jk,j}$, while $\mathbf D_{jk}$ is the variance of channel estimation error.

\section{Stochastic Matrix FP Methods}
\label{sec:sfp}

The matrix FP has been extensively used for the static precoding task (i.e., when the channels $\mathbf{H}_{jk,j}$ are all fixed and known precisely). It is natural to extend the existing matrix FP methods for the stochastic precoding case, as discussed in the first part of this section. Nevertheless, the second part of the section shows that applying the matrix FP to the matrix ratios directly inside expectation cannot lead to efficient optimization, because the auxiliary variables introduced by the matrix FP are difficult to decide. As such, we further construct a lower bound on the original objective function and use it to approximate the problem.


\subsection{Stochastic Extension of  Matrix FP}
\label{subsec:sqt}

We start by showing how the quadratic transform and the Lagrangian dual transform in \cite{shen_d2d} can be generalized for the stochastic case. To avoid complication in notation, and to present the results in full generality, we consider the following generic problem formulation. Let $\mathbf{A}_n(x\mid\theta^+_n)\in\mathbb S^{M\times M}_{+}$ and $\mathbf{B}_n(x\mid\theta^-_n)\in\mathbb S^{M\times M}_{++}$ be a pair of matrix-valued functions of $x$, for $n=1,2,\ldots,N$, where $\{\theta^+_n,\theta^-_n\}$ is a set of implicit random variables that impact the function outputs. To ease notation, the arguments $(x\mid\theta^+_n)$ and $(x\mid\theta^-_n)$ are omitted in the rest of this section, unless we wish to emphasize their specific values. Let us consider an abstraction of the stochastic precoding problem:
\begin{align}
\label{log prob}
    \underset{x\in\mathcal X}{\text{maximize}}&\quad\mathbb{E}\Bigg[\sum^N_{n=1}\omega_n\log\Big|\mathbf{I}+\mathbf{A}_n\mathbf B_n^{-1}\Big|\Bigg],
\end{align}
where $\mathcal X$ is a nonempty constraint set and each $\omega_n\ge0$ is a positive weight. Following the Lagrangian dual transform \cite{shen_d2d}, we try to move the matrix ratio term out of the log-det function, as stated in the following theorem. 
\begin{theorem}
\label{thm:Lagrange}
Problem \eqref{log prob} is equivalent to
\begin{align}
    &\underset{x\in\mathcal X}{\text{maximize}}\; \mathbb{E}\bigg[\sup_{\bm\Gamma}\sum^N_{n=1}\omega_n\Big(\log|\mathbf I+\bm\Gamma_n|-\mathrm{Tr}\big(\bm\Gamma_n\big)\notag\\
    &\qquad\quad+\mathrm{Tr}\big((\mathbf I+\bm\Gamma_n)\sqrt{\mathbf{A}_n}^H(\mathbf{A}_n+\mathbf{B}_n)^{-1}\sqrt{\mathbf{A}_n}\big)\Big)\bigg]\label{prob:sqt}
\end{align}
in the sense that $x$ is a global solution (or a stationary point solution) to \eqref{log prob} iff it is a global solution (or a stationary point solution) to \eqref{prob:sqt},
where a positive semidefinite auxiliary variable $\bm\Gamma_n\in\mathbb S^{M\times M}_+$ is introduced for each matrix ratio.
\end{theorem}
\begin{IEEEproof}
    For any feasible $x\in\mathcal X$, the optimal $\bm\Gamma_n$ in \eqref{prob:sqt} is given by 
\begin{equation}
\label{opt gamma}
    \bm\Gamma^\star_n = \sqrt{\mathbf A_n}^H\mathbf B_n^{-1}\sqrt{\mathbf A_n}.
\end{equation}
Substituting $\bm\Gamma^\star_n$ into problem \eqref{prob:sqt} recovers problem \eqref{log prob}.
\end{IEEEproof}

When $\bm\Gamma$ is fixed, we only need to look at the second line of \eqref{prob:sqt} for optimizing $x$. This subproblem boils down to
\begin{align}\label{prob:esfmr}
    \underset{x\in\mathcal X}{\text{maximize}}&\quad\mathbb{E}\Bigg[\sum^N_{n=1}\mathrm{Tr}\big(\mathbf{Q}_n\sqrt{\mathbf A_n}^H \widetilde{\mathbf B}_n^{-1}\sqrt{\mathbf A_n}\big)\Bigg],
\end{align}
where $\mathbf Q_n\in\mathbb S^{M\times M}_{++}$ is a positive definite matrix, and $\widetilde{\mathbf{B}}_n\in\mathbb S^{M\times M}_{++}$ is a random matrix-valued function of $x$. We now mimic the quadratic transform in \cite{shen_d2d}, by decoupling each matrix ratio, as stated in the following theorem.
\begin{theorem}
\label{thm:QT}
Problem \eqref{prob:esfmr} is equivalent to
\begin{align}
 \underset{x\in\mathcal X}{\text{maximize}}&\quad \mathbb{E}\bigg[\sup_{\mathbf Y}\sum^N_{n=1}\mathrm{Tr}\Big(\mathbf Q_n\big(\sqrt{\mathbf A_n}^H\mathbf Y_n\notag\\
 &\qquad\qquad+\mathbf Y_n^H\sqrt{\mathbf A_n}-\mathbf Y_n^H\widetilde{\mathbf B}_n\mathbf Y_n\big)\Big)\bigg]
 \label{prob:QT new}
\end{align}   
\end{theorem}
in the sense that $x$ is a global solution (or a stationary point solution) to \eqref{prob:esfmr} iff it is a global solution (or a stationary point solution) to \eqref{prob:QT new},
where the auxiliary variable $\mathbf Y_n\in\mathbb C^{M\times M}$ is introduced for each matrix ratio.
\begin{IEEEproof}
For any feasible $x\in\mathcal X$, the optimal $\mathbf Y_n$ in \eqref{prob:QT new} is given by
\begin{equation}
\mathbf{Y}_n^\star=\Big(\widetilde{\mathbf{B}}_n(x\mid\theta^+_n,\theta^-_n)\Big)^{-1}\sqrt{\mathbf{A}_n(x\mid\theta^+_n)}.
\end{equation}
Substituting $\mathbf{Y}_n^\star$ into problem \eqref{prob:QT new} recovers problem \eqref{prob:esfmr}.
\end{IEEEproof}
However, Theorem \ref{thm:Lagrange} and Theorem \ref{thm:QT} alone are insufficient for solving the stochastic beamforming problem, because their auxiliary variables are difficult to decide. To address this issue, we further propose a lower bound in Section \ref{logr_lb}, which is a key result of this paper.

\subsection{Proposed Lower Bound}
\label{logr_lb}

Combining the transformations in the above two theorems, we convert problem \eqref{log prob} to
\begin{align}
    \underset{x\in\mathcal X}{\text{maximize}}&\quad f(x):= \mathbb E\bigg[\sup_{\bm\Gamma,\mathbf Y}g(
x,\bm\Gamma,\mathbf Y\mid\theta^+,\theta^-)\bigg],
    \label{new prob}
\end{align}
where
\begin{align}
g(x,\bm\Gamma,\mathbf Y&\mid\theta^+,\theta^-)=\sum^N_{n=1}\omega_n\Big[\log|\mathbf I+\bm\Gamma_n|-\mathrm{Tr}(\bm\Gamma_n)\notag\\
&\quad+\mathrm{Tr}\big((\mathbf I+\bm\Gamma_n)(\sqrt{\mathbf{A}_n}^H\mathbf{Y}_n+\mathbf{Y}_n^H\sqrt{\mathbf{A}_n})\big)\notag\\
&\quad-\mathrm{Tr}\big((\mathbf I+\bm\Gamma_n)\mathbf{Y}_n^H(\mathbf{A}_n+\mathbf{B}_n)\mathbf{Y}_n\big)\Big]
\end{align}
with the auxiliary variables $\mathbf Y_n\in\mathbb C^{M\times M}$ and $\bm\Gamma_n\in\mathbb S^{M\times M}_+$.

It is then tempting to optimize $(x,\bm\Gamma,\mathbf Y)$ in the above new problem in an iterative fashion, just as in the static precoding problem case. But we point out a subtle issue with this method in the following. When $x$ and $\bm\Gamma$ are fixed, each $\mathbf Y_n$ is optimally determined as
\begin{equation}
\label{opt Y}
\mathbf{Y}_n^\star=\Big(\mathbf A_n(x\mid\theta^+_n)+\mathbf{B}_n(x\mid\theta^-_n)\Big)^{-1}\sqrt{\mathbf{A}_n(x\mid\theta^+_n)}.
\end{equation} 
When $x$ and $\mathbf Y$ are fixed, each $\bm \Gamma_n$ is optimally determined as
\begin{equation}
\label{}
    \bm\Gamma^\star_n = \sqrt{\mathbf A_n(x\mid\theta^+_n)}^H\Big(\mathbf B_n(x\mid\theta^-_n)\Big)^{-1}\sqrt{\mathbf A_n(x\mid\theta^+_n)}.
\end{equation} 
Note that the optimal updates of $\mathbf Y_n$ and $\bm\Gamma_n$ depend on the current realizations of the random variables $\{\theta^+_n,\theta^-_n\}$, which are not available in our problem case. As a consequence, the conventional way of using FP does not work for stochastic precoding.

Like the precoding variable, the auxiliary variables $\mathbf Y$ and $\bm\Gamma$ ought to depend on the channel statistics rather than a particular realization. To remedy the problem, we propose interchanging $\mathbb E[\cdot]$ and $\sup_{\bm\Gamma,\mathbf Y}\{\cdot\}$ in \eqref{new prob}; subsequently, $\sup_{\bm\Gamma,\mathbf Y}$ can be further incorporated into ``maximize'' as
\begin{subequations}\label{new prob 2}
\begin{align}
\underset{x\in\mathcal X,\mathbf \Gamma,\mathbf Y}{\text{maximize}}&\quad \hat f(x,\bm\Gamma,\mathbf Y):= \mathbb E\bigg[g(
x,\mathbf Y,\bm\Gamma\mid\theta^+_n,\theta^-_n)\bigg],\\
\text{subject to}& \quad \bm\Gamma_{n}\in\mathbb S^{M\times M}_+,\; \forall n, \\
&\quad \mathbf Y_{n}\in\mathbb C^{M\times M},\; \forall n.
\end{align}
\end{subequations}
It can be seen that the optimal updates of $\mathbf Y_n$ and $\bm\Gamma_n$ are now independent of the realizations of $\{\theta^+_n,\theta^-_n\}$.

We now justify the above problem transformation by showing that the new problem is in essence to maximize a lower bound of the original objective function in \eqref{log prob}.

\begin{theorem}
\label{thm:lower bound}
For any feasible tuple $(x,\bm\Gamma,\mathbf Y)$, we have
\begin{equation}
\label{lower bound 1}
\hat f(x,\bm\Gamma,\mathbf Y)\le\mathbb{E}\Bigg[\sum^N_{n=1}\omega_n\log\Big|\mathbf{I}+\mathbf{A}_n\mathbf B_n^{-1}\Big|\Bigg].
\end{equation}
\end{theorem}
\begin{IEEEproof}
We first show that the interchange of $\mathbb E$ and $\sup$ in \eqref{lower bound 1} is mathematically well-posed. Since the feasible set of $(\bm\Gamma,\mathbf Y)$ needs to be nonempty, and that the objective function $g(x,\bm\Gamma,\mathbf Y\mid\theta^+,\theta^-)$ is upper bounded, so $\sup_{\mathbf \Gamma,\mathbf Y} g(x,\bm\Gamma,\mathbf Y\mid\theta^+,\theta^-)$ must exist. Note also that $g(x,\bm\Gamma,\mathbf Y\mid\theta^+,\theta^-)$ is a continuous function defined on the measurable domain $(x,\mathbf \Gamma,\mathbf Y)\in \mathcal X\times \mathbb S_+^{M\times M}\times \mathbb C^{M\times M}$, so $g(x,\bm\Gamma,\mathbf Y\mid\theta^+,\theta^-)$ is a measurable function, and its supremum is a measurable function as well. As a result, $g(x,\bm\Gamma,\mathbf Y\mid\theta^+,\theta^-)$ and $\sup_{\mathbf \Gamma,\mathbf Y} g(x,\bm\Gamma,\mathbf Y\mid\theta^+,\theta^-)$ are both Lebesgue integrable, so $\mathbb{E}[\sup_{\mathbf \Gamma,\mathbf Y} g(x,\bm\Gamma,\mathbf Y\mid\theta^+,\theta^-)]$ and $\mathbb{E}[g(x,\bm\Gamma,\mathbf Y\mid\theta^+,\theta^-)]$ must exist. Further, since $\mathcal X$ is nonempty, $\sup_{\mathbf \Gamma,\mathbf Y}\mathbb{E}[g(x,\bm\Gamma,\mathbf Y\mid\theta^+,\theta^-)]$ must exist.

Because of the equivalence between \eqref{log prob} and \eqref{new prob}, it suffices to show that
\begin{equation}
    \hat f(x,\bm\Gamma,\mathbf Y) \le f(x).
\end{equation}
We then have
\begin{subequations}
    \begin{align}
\hat f(x,\bm\Gamma,\mathbf Y) &\le \sup_{\bm\Gamma,\mathbf Y}\hat f(x,\bm\Gamma,\mathbf Y)\\
&= \sup_{\bm\Gamma,\mathbf Y}\mathbb E[g(
x,\bm\Gamma,\mathbf Y\mid\theta^+_n,\theta^-_n)]\\
&\le \mathbb E\bigg[\sup_{\bm\Gamma,\mathbf Y}g(
x,\bm\Gamma,\mathbf Y\mid\theta^+_n,\theta^-_n)\bigg]\label{ieq:lb}\\
&= f(x).
\end{align}
\end{subequations}
The proof is complete.
\end{IEEEproof}
\begin{figure*}[t]
\begin{multline}
\hat f(\mathbf{V},\mathbf{\Gamma},\mathbf{Y})= \sum_{(j,k)}\omega_{jk}\Bigg[\log\big|\mathbf{I}+\mathbf{\Gamma}_{jk}\big|-\mathrm{Tr}\big(\mathbf{\Gamma}_{jk}\big)+\mathrm{Tr}\Bigg(\big(\mathbf{I}+\mathbf{\Gamma}_{jk}\big)\Bigg(2\Re\big\{\mathbf{V}_{jk}^H\mathbf{C}_{jk}^H\mathbf{Y}_{jk}\big\}-\mathbf{Y}_{jk}^H\Bigg(\sum_{(\ell,s)}\mathbf{U}_{jk,\ell s}+\sigma^2\mathbf{I}\Bigg)\mathbf{Y}_{jk}\Bigg)\Bigg)\Bigg] \tag{26}
\label{obj:Epgy}
\end{multline}
\hrule
\end{figure*}

In problem \eqref{new prob 2}, to make the bound $\hat f(x,\bm\Gamma,\mathbf Y)$ as tight as possible, we maximize $\hat f(x,\bm\Gamma,\mathbf Y)$ over $\mathbf V$ and $\bm\Gamma$ to get:
\begin{align}
    \tilde{\bm\Gamma}^\star_n &= \mathbb E[\sqrt{\mathbf A_n}]^H\Big(\mathbb E[\mathbf A_n]+\mathbb E[\mathbf B_n]\nonumber\\
        &\qquad-\mathbb E[\sqrt{\mathbf A_n}]\mathbb E[\sqrt{\mathbf A_n}]^H\Big)^{-1}\mathbb E[\sqrt{\mathbf A_n}],\\
        \tilde{\mathbf{Y}}_n^\star&=\Big(\mathbb E[\mathbf A_n]+\mathbb E[\mathbf B_n]\Big)^{-1}\mathbb E[\sqrt{\mathbf A_n}],
\end{align}
where the arguments of $\mathbf A_n$ and $\mathbf B_n$ are omitted to ease notation. Importantly, the above two solutions can be treated as the functions of $x$, so substituting them into $\hat f(x,\bm\Gamma,\mathbf Y)$ leads to a lower bound $\tilde f(x)$ without $\bm\Gamma$ and $\mathbf Y$, i.e.,
\begin{equation}\label{eq:lower_boundftx}
    f(x) \ge \tilde f(x) := \hat f(x, \tilde{\mathbf \Gamma}^\star,\tilde{\mathbf Y}^\star),
\end{equation}
where the equality holds when $\mathrm{Pr}\{(\theta^+_n,\theta^-_n):\mathbf \Gamma^\star_n=\tilde{\bm\Gamma}^\star_n\;\text{and}\;\mathbf{Y}_n^\star=\tilde{\mathbf Y}^\star_n\}=1$, i.e., when $(\mathbf \Gamma,\mathbf Y)$ are deterministic functions of moments. As compared to the primal objective $f(x)$, the new bound $\tilde f(x)$ looks simpler in that it gets rid of the expectation. However, $\tilde f(x)$ is still difficult to optimize directly because of its nonconvexity. If we consider the parameterized bound $\hat f(x,\bm\Gamma,\mathbf Y)$ instead, then the solution of $x$ can be readily obtained in closed form with the auxiliary variables $(\bm\Gamma,\mathbf Y)$ updated iteratively; this can be recognized as a minorization-maximization (MM) procedure with provable performance.

In the next section, we show that the new problem \eqref{new prob 2} allows the precoding variable and the auxiliary variables to be efficiently optimized based on the first and second moments of fading channels.


\section{FP-Based Stochastic Precoding}
\label{sec:alg}

We now specialize the above matrix FP method for the stochastic precoding problem \eqref{prob:orig}. In light of the following correspondences:
\begin{align}
    \sqrt{\mathbf A_n(x\mid\theta^+_n)} &\;\Longleftrightarrow\; \mathbf H_{jk,j}\mathbf V_{jk},\\
    \mathbf B_n(x\mid\theta^-_n) &\;\Longleftrightarrow\; \mathbf F_{jk},
\end{align}
problem \eqref{prob:orig} is converted to 
\begin{subequations}\label{prob:bound}
    \begin{align}
        \underset{\mathbf{V},\bm\Gamma,\mathbf Y}{\text{maximize}}&\quad \hat f(\mathbf{V},\mathbf{\Gamma},\mathbf{Y})\\
    \text{subject to}& \quad \sum^K_{k=1}\text{Tr}(\mathbf{V}_{jk}\mathbf{V}_{jk}^H)\leq P,\; \forall j\label{cons:power}\\
    &\quad\, \bm\Gamma_{jk}\in\mathbb S^{M_r\times M_r}_+,\; \forall (j,k) \label{cons:gamma}\\
    &\quad\, \mathbf Y_{jk}\in\mathbb C^{M_r\times M_r},\; \forall (j,k),\label{cons:y}
    \end{align}
\end{subequations}
where the new objective function $\hat f(\mathbf{V},\mathbf{\Gamma},\mathbf{Y})$ is shown in \eqref{obj:Epgy}, with the shorthand
\setcounter{equation}{26}
\begin{align}
\mathbf U_{jk,\ell s} = \mathbb E\big[\mathbf{H}_{jk,\ell}\mathbf{V}_{\ell s}\mathbf{V}_{\ell s}^H\mathbf{H}_{jk,\ell}^H\big].
\end{align}
Each entry of $\mathbf U_{jk,\ell s}\in\mathbb S^{M_r\times M_r}_+$ can be obtained from the current $\mathbf V$ and the second moment $\mathbf D_{jk,\ell}$ as
\allowdisplaybreaks
\begin{align}
&\big(\mathbf U_{jk,\ell s}\big)_{m,n}\notag\\
&=\sum_{m'=1}^{M_t}\sum_{n'=1}^{M_t}\big(\mathbf V_{\ell s}\mathbf V^H_{\ell s}\big)_{m',n'}\mathbb{E}\big[\big(\mathbf{H}_{jk,\ell}\big)_{m,m'}\big(\mathbf{H}^c_{jk,\ell}\big)_{n,n'}\big]\notag\\
&= \sum_{m'=1}^{M_t}\sum_{n'=1}^{M_t}\Big(\big(\mathbf V_{\ell s}\mathbf V^H_{\ell s}\big)_{m',n'}\notag\\
&\qquad\times(\mathbf D_{jk,\ell})_{(m'-1)M_r+m,(n'-1)M_r+n}\Big)
\label{U}
\end{align}
We then optimize $(\mathbf{V},\bm\Gamma,\mathbf Y)$ in an iterative fashion. When $\mathbf{V}$ and $\mathbf Y$ are both held fixed, each $\bm\Gamma_{jk}$ can be optimally determined as
\begin{multline}\label{eq:Ga_opt}
\mathbf{\Gamma}_{jk}^\star=\mathbf{V}_{jk}^H\mathbf{C}_{jk}^H\Bigg(\sigma^2\mathbf{I}+\sum_{(\ell,s)}\mathbf U_{jk,\ell s}\\-\mathbf{C}_{jk}\mathbf{V}_{jk}\mathbf{V}_{jk}^H\mathbf{C}_{jk}^H\Bigg)^{-1}\mathbf{C}_{jk}\mathbf{V}_{jk}.
\end{multline}
When $\mathbf{V}$ and $\bm\Gamma$ are both held fixed, each $\mathbf Y_{jk}$ can be optimally determined as
\begin{equation}\label{eq:Y_opt}
\mathbf{Y}_{jk}^\star=\Bigg(\sigma^2\mathbf{I}+\sum_{(\ell,s)}\mathbf U_{jk,\ell s}\Bigg)^{-1}\mathbf{C}_{jk}\mathbf{V}_{jk}.
\end{equation}
After $\bm\Gamma$ has been optimized according to \eqref{eq:Ga_opt} for the current $\mathbf Y$, the solution of each $\mathbf V_{jk}$ is given by
\begin{equation}\label{eq:upPbisearch}
\mathbf{V}_{jk}=\Bigg(\eta_j\mathbf{I}+\sum_{(\ell,s)}\omega_{\ell s}\bm \Lambda_{j,\ell s}\Bigg)^{-1}\omega_{jk}\mathbf{C}_{jk}^H\mathbf{Y}_{jk}(\mathbf{I}+\mathbf{\Gamma}_{jk}),
\end{equation}
where the Lagrangian multiplier $\eta_j\ge0$ for the power constraint is optimally determined as
\begin{align}
\label{Lagrange multiplier}
    \eta_j^\star=\min\Bigg\{\eta_j\geq0:\sum^K_{k=1}\mathrm{Tr}(\mathbf{V}_{jk}\mathbf{V}_{jk}^H)\leq P\Bigg\},
\end{align}
and the matrix $\bm \Lambda_{j,\ell s}\in\mathbb S^{M_t\times M_t}_{+}$ is given by
\begin{equation}
\bm \Lambda_{j,\ell s} = \mathbb{E}\big[\mathbf{H}_{\ell s,j}^H\mathbf{Y}_{\ell s}(\mathbf{I}+\mathbf{\Gamma}_{\ell s})\mathbf{Y}_{\ell s}^H\mathbf{H}_{\ell s,j}\big].
\end{equation}
We can use $\mathbf D_{jk,\ell}$ to recover each entry of $\bm \Lambda_{j,\ell s}$ as
\begin{align}
&\big(\bm \Lambda_{j,\ell s}\big)_{m,n}\notag\\
&=\sum_{m'=1}^{M_r}\sum_{n'=1}^{M_r}\Big(\big(\mathbf{Y}_{\ell s}(\mathbf{I}+\mathbf{\Gamma}_{\ell s})\mathbf{Y}_{\ell s}^H\big)_{m',n'}\notag\\
&\qquad\qquad\qquad\quad\times\mathbb{E}\big[\big(\mathbf{H}^c_{\ell s,j}\big)_{m',m}\big(\mathbf{H}_{\ell s,j}\big)_{n',n}\big]\Big)\notag\\
&= \sum_{m'=1}^{M_r}\sum_{n'=1}^{M_r}\Big(\big(\mathbf{Y}_{\ell s}(\mathbf{I}+\mathbf{\Gamma}_{\ell s})\mathbf{Y}_{\ell s}^H\big)_{m',n'}\notag\\
&\qquad\qquad\qquad\times(\mathbf D_{\ell s,j}^c)_{(m-1)M_r+m',(n-1)M_r+n'}\Big).
\label{Lambda}
\end{align}

Algorithm \ref{alg1} summarizes the above steps for stochastic precoding. Its per-iteration complexity equals $\mathcal{O}(L^2K^2M_r^2M_t^2+LKM_r^3+LM_t^3)$. The following theorem provides a convergence analysis for this algorithm.
\begin{algorithm}[t]
    \caption{Stochastic Precoding Based on Matrix FP}
    \label{alg1}
    \begin{algorithmic}[1]
        \STATE \textbf{input:} first and second moments $\{\mathbf C,\mathbf D\}$ of fading channels
        \REPEAT
        \STATE compute $\mathbf{U}$ by \eqref{U} and then update $\mathbf{Y}$ by \eqref{eq:Y_opt}
        \STATE update $\bm{\Gamma}$ by \eqref{eq:Ga_opt}
        \STATE compute $\bm\Lambda$ by \eqref{Lambda}
        and then update $\mathbf{V}$ by \eqref{eq:upPbisearch}
        \UNTIL{the objective value converges}
        \STATE \textbf{output:} precoding matrix $\mathbf V$
    \end{algorithmic}
\end{algorithm}

\begin{proposition}
\label{prop:stationary point}
    Algorithm \ref{alg1} guarantees convergence to a stationary point $(\mathbf V,\bm\Gamma,\mathbf Y)$ of problem \eqref{prob:bound}. Moreover, the new objective in \eqref{prob:bound} is a lower bound to the original objective:
    \begin{equation}
        \hat f(\mathbf{V},\mathbf{\Gamma},\mathbf{Y}) \le \sum^L_{j=1}\sum^K_{k=1}\mathbb{E}\Big[\omega_{jk}\mathcal{R}_{jk}\Big].
    \end{equation}
\end{proposition}
\begin{IEEEproof}
    The convergence result follows by the property of FP as shown in \cite{fp_ii}. The lower bound result follows by Theorem \ref{thm:lower bound}.
\end{IEEEproof}

The following remarks in order point out some subtle aspects of Algorithm \ref{alg1} and Proposition \ref{prop:stationary point}.
\begin{remark}
    The proposed algorithm guarantees a stationary point of the lower bound but not the original problem. We also wish to point out that the above result is by no means trivial. In the existing literature, \cite{aal} show that their algorithms converge to stationary points of the new problems, but the connection between the new problems and the original problem cannot be proved. Moreover, \cite{gaussianerror4, shi21, gaussianerror3} use the upper bound to approximate the problem and show that their solutions are the stationary points of the new problem. However, since the original problem is a maximization problem, it makes more sense to consider a lower bound as in our work. To the best of our knowledge, \cite{swmmse} and \cite{spwmmse} are the only works that guarantee stationary points of the original problem, but they are in essence based on Monte Carlo method and require a huge set of instantaneous CSI for the training purpose.
\end{remark}
\begin{remark}
    The proposed method is in essence to approximate the fading channels as Gaussian. After all, with only the first-order and second-order moments available, it is most reasonable to adopt the Gaussian approximation. Nevertheless, it is not that trivial to show that the Gaussian approximation can yield a lower bound for this particular problem case.
\end{remark}
\begin{remark}
The computations of $\mathbf U$ and $\bm\Lambda$ in Algorithm \ref{alg1} can be further simplified when fading channels are known to be Rayleigh or Rician. In that case, each $\mathbf H_{jk,\ell}$ has a Gaussian distribution:
\begin{equation}
    \mathbf{H}_{jk,\ell} = \widetilde{\mathbf{H}}_{jk,\ell}+\mathbf{W}_{jk,\ell}\odot\mathbf{X}_{jk,\ell},
\end{equation}
where $\widetilde{\mathbf{H}}_{jk,\ell}\in\mathbb C^{M_r\times M_t}$ and $\mathbf{W}_{jk,\ell}\in\mathbb C^{M_r\times M_t}$ are fixed matrices, and $\mathbf{X}_{jk,\ell}\in\mathbb C^{M_r\times M_t}$ is a random matrix with each entry drawn i.i.d. from the standard complex Gaussian distribution $\mathcal{CN}(0,1)$. Then $\mathbf U$ and $\bm\Lambda$ are directly given by
\begin{align}
    &\mathbf U_{jk,\ell s} = \widetilde{\mathbf{H}}_{jk,\ell}\mathbf{V}_{\ell s}\mathbf{V}_{\ell s}^H\widetilde{\mathbf{H}}_{jk,\ell}^H\notag\\
    &\;+\mathrm{diag}\Big(\big(\mathbf{W}_{jk,\ell}\odot\mathbf{W}_{jk,\ell}\big)\mathrm{Diag}\big(\mathbf{V}_{\ell s}\mathbf{V}_{\ell s}^H\big)\Big),\\
    &\bm \Lambda_{j,\ell s} = \widetilde{\mathbf{H}}_{jk,\ell}^H\mathbf{Y}_{\ell s}(\mathbf{I}+\mathbf{\Gamma}_{\ell s})\mathbf{Y}_{\ell s}^H\widetilde{\mathbf{H}}_{jk,\ell}\notag\\
    &\;+\mathrm{diag}\Big(\big(\mathbf{W}_{jk,\ell}^H\odot\mathbf{W}_{jk,\ell}^H\big)\mathrm{Diag}\big(\mathbf{Y}_{\ell s}(\mathbf{I}+\mathbf{\Gamma}_{\ell s})\mathbf{Y}_{\ell s}^H\big)\Big),
\end{align}
so we no longer need to compute these matrices entry by entry as in \eqref{U} and \eqref{Lambda}. As a result, the per-iteration complexity of Algorithm \ref{alg1} reduces to $\mathcal{O}(L^2K^2M_rM_t^2+L^2K^2M_r^2M_t+LKM_r^3+LM_t^3)$.
In particular, if we further let $M_r=1$ then Algorithm \ref{alg1} reduces to the precoding method in \cite{shi21}. 
\end{remark}

\section{Acceleration for Large-Scale MIMO}
\label{sec:fast FP}
\begin{figure*}
\begin{multline}
\label{obj:Epgy new}
\hat f(\mathbf{V},\mathbf{\Gamma},\mathbf{Y})=\sum_{(j,k)}\Bigg(\omega_{jk}\log\big|\mathbf{I}+\mathbf{\Gamma}_{jk}\big|-\omega_{jk}\mathrm{Tr}\big(\mathbf{\Gamma}_{jk}\big)\\
+\mathrm{Tr}\Bigg(2\omega_{jk}\big(\mathbf{I}+\mathbf{\Gamma}_{jk}\big)\Re\big\{\mathbf{V}_{jk}^H\mathbf{C}_{jk}^H\mathbf{Y}_{jk}\big\}-\omega_{jk}\sigma^2\big(\mathbf{I}+\mathbf{\Gamma}_{jk}\big)\mathbf{Y}_{jk}^H\mathbf{Y}_{jk}-\mathbf{V}_{jk}^H\Bigg(\sum_{(\ell,s)}\omega_{\ell s}\mathbf{\Lambda}_{j,\ell s}\Bigg)\mathbf{V}_{jk}\Bigg)\Bigg)\tag{40}
\end{multline}
\hrule
\begin{multline}\label{obj:pgyz}
\zeta(\mathbf{V},\mathbf{\Gamma},\mathbf{Y},\mathbf{Z})
=\sum_{(j,k)}\bigg(\omega_{jk}\log|\mathbf{I}+\mathbf{\Gamma}_{jk}|-\omega_{jk}\mathrm{Tr}(\mathbf{\Gamma}_{jk})+2\mathrm{Tr}\Big(\Re\Big\{\omega_{jk}(\mathbf{I}+\mathbf{\Gamma}_{jk})\mathbf{V}_{jk}^H\mathbf{C}_{jk}^H\mathbf{Y}_{jk}+\mathbf{V}_{jk}^H(\alpha_{j}\mathbf{I}-\bm\Xi_j)\mathbf{Z}_{jk}\Big\}\Big)\\
+\mathrm{Tr}\Big(\mathbf{Z}_{jk}^H(\bm\Xi_j-\alpha_{j}\mathbf{I})\mathbf{Z}_{jk}-\alpha_{j}\mathbf{V}_{jk}^H\mathbf{V}_{jk}-\omega_{jk}\sigma^2(\mathbf{I}+\mathbf{\Gamma}_{jk})\mathbf{Y}_{jk}^H\mathbf{Y}_{jk}\Big)\bigg)\tag{42}
\end{multline}
\hrule
\end{figure*}
In this section we cope with such a challenge: for the large-scale MIMO case with $M_t$ being a large number, the $M_t\times M_t$ matrix inverse in \eqref{eq:upPbisearch} renders the update of $\mathbf V$ computationally formidable in Step 5 of Algorithm \ref{alg1}. The main idea is to get rid of this $M_t\times M_t$ matrix inverse by further modifying the matrix FP. We start by introducing a useful inequality from \cite{chenfastfp,nonhomobound}.



\begin{lemma}\label{lemma:nonhomo}
Consider any two ${M\times M}$ Hermitian matrices $\mathbf{L},\mathbf{K}$ satisfying $\mathbf{L}\preceq\mathbf{K}$. For any $\mathbf{X},\mathbf{Z}\in\mathbb{C}^{M\times N}$, we have
\begin{multline}
\label{nonhomo inequality}
\mathrm{Tr}\big(\mathbf{X}^H\mathbf{L}\mathbf{X}\big)\leq\mathrm{Tr}\big(\mathbf{X}^H\mathbf{K}\mathbf{X}+2\Re\{\mathbf{X}^H(\mathbf{L}-\mathbf{K})\mathbf{Z}\}\\+\mathbf{Z}^H(\mathbf{K}-\mathbf{L})\mathbf{Z}\big),
\end{multline}
where the equality holds if $\mathbf{Z}=\mathbf{X}$.
\end{lemma}

To see how the above inequality may assist with our problem case, we first rewrite $\hat f(\mathbf{V},\mathbf{\Gamma},\mathbf{Y})$ in \eqref{obj:Epgy} in a quadratic form of $\mathbf V_{jk}$, as shown in \eqref{obj:Epgy new}. Now observe that the $M_t\times M_t$ matrix inverse in \eqref{eq:upPbisearch} is caused by the quadratic term $\mathbf{V}_{jk}^H(\sum_{(\ell,s)}\omega_{\ell s}\mathbf{\Lambda}_{j,\ell s})\mathbf{V}_{jk}$ in \eqref{obj:Epgy new}. Let us treat $\sum_{(\ell,s)}\omega_{\ell s}\mathbf{\Lambda}_{j,\ell s}$ as $\mathbf L$ in \eqref{nonhomo inequality}. The main idea here is to replace this matrix with a ``simpler'' matrix $\mathbf K$ in the quadratic term by using Lemma \ref{lemma:nonhomo};
since we seek to eliminate matrix inverse, it is desirable to let $\mathbf K=\alpha \mathbf I$ so that its inverse is directly given by $\mathbf K^{-1}=(1/\alpha) \mathbf I$. But how do we choose the factor $\alpha$ to meet the condition $\mathbf{L}\preceq\mathbf{K}$ in Lemma \ref{lemma:nonhomo}? As suggested in \cite{shen_accqt}, we may set $\alpha$ to the largest eigenvalue of $\mathbf L$.

\begin{algorithm}[t]
    \caption{Fast Stochastic Precoding for Large-Scale MIMO}
    \label{alg2}
    \begin{algorithmic}[1]
        \STATE \textbf{input:} first and second moments $\{\mathbf C,\mathbf D\}$ of fading channels
        \REPEAT
        \STATE compute $\mathbf{U}$ by \eqref{U} and then update $\mathbf{Y}$ by \eqref{eq:Y_opt}
        \STATE update $\bm{\Gamma}$ by \eqref{eq:Ga_opt}
        \STATE update $\mathbf{Z}$ by \eqref{eq:z_opt}
        \STATE compute $\bm\Lambda$ by \eqref{Lambda}
        and then update $\mathbf{V}$ by \eqref{eq:upPfast}
        \UNTIL{the objective value converges}
        \STATE \textbf{output:} precoding matrix $\mathbf V$
    \end{algorithmic}
\end{algorithm}

By applying the inequality \eqref{nonhomo inequality} to $\hat f(\mathbf{V},\mathbf{\Gamma},\mathbf{Y})$ in \eqref{obj:Epgy}, we obtain a lower bound 
\setcounter{equation}{40}
\begin{equation}
\zeta(\mathbf{V},\mathbf{\Gamma},\mathbf{Y},\mathbf Z)\le \hat f(\mathbf{V},\mathbf{\Gamma},\mathbf{Y}),
\end{equation}
which is shown in \eqref{obj:pgyz} with the shorthands
\setcounter{equation}{42}
\begin{equation}
\bm\Xi_j=\sum_{(\ell,s)}\omega_{\ell s}\mathbf{\Lambda}_{j,\ell s}
\end{equation}
and
\begin{equation}
    \alpha_j=\lambda_{\max}(\bm\Xi_j).
\end{equation}
\begin{remark}
The gap between $\hat f$ and $\zeta$ can be computed as
\begin{subequations}\label{eq:fxigap}
    \begin{align}
        \hat f - \zeta =& \sum_{(j,k)}\mathrm{Tr}\Big( (\mathbf{V}_{jk}-\mathbf{Z}_{jk})^H\nonumber \\
        &\qquad\times\left( \lambda_{\max}(\mathbf{\Xi}_j)\mathbf{I} - \mathbf{\Xi}_j \right) (\mathbf{V}_{jk}-\mathbf{Z}_{jk}) \Big)\\
        \leq&\sum_{(j,k)}\lambda_{\max}\left( \lambda_{\max}(\mathbf{\Xi}_j)\mathbf{I} - \mathbf{\Xi}_j \right)\nonumber\\
        &\qquad\times \mathrm{Tr}\Big( (\mathbf{V}_{jk}-\mathbf{Z}_{jk})^H(\mathbf{V}_{jk}-\mathbf{Z}_{jk}) \Big)\\
        =&\sum_{(j,k)}\Big(\lambda_{\max}(\mathbf{\Xi}_j)-\lambda_{\min}(\mathbf{\Xi}_j)\Big)\nonumber\\
        &\qquad\times \mathrm{Tr}\Big( (\mathbf{V}_{jk}-\mathbf{Z}_{jk})^H(\mathbf{V}_{jk}-\mathbf{Z}_{jk}) \Big).
    \end{align}
\end{subequations}
Observe that the gap increases with the spectral spread of $\Xi_j$.
\end{remark}

Problem \eqref{prob:bound} is now turned to
\begin{subequations}\label{prob:fastbound}
    \begin{align}
        \underset{\mathbf{V},\mathbf Z,\bm\Gamma,\mathbf Y}{\text{maximize}}&\quad \zeta(\mathbf{V},\mathbf{Z},\mathbf{\Gamma},\mathbf{Y})\\
    \text{subject to}& \quad \eqref{cons:power},\eqref{cons:gamma},\eqref{cons:y}\\
    &\quad\, \mathbf Z_{jk}\in\mathbb C^{M_t\times M_r},\; \forall (j,k).
    \end{align}
\end{subequations}
As before, we optimize these variables in an iterative fashion. When $(\mathbf{V},\mathbf{\Gamma},\mathbf{Y})$ are fixed, as shown in Lemma \ref{lemma:nonhomo}, each $\mathbf Z_j$ can be optimally determined as
\begin{align}\label{eq:z_opt}
    \mathbf{Z}_{jk}^\star = \mathbf{V}_{jk},
\end{align}
The iterative optimal updates of $\mathbf Y$ and $\bm \Gamma$ are still given by \eqref{eq:Y_opt} and \eqref{eq:Ga_opt}, respectively. Moreover, for fixed $(\mathbf{Y},\mathbf{\Gamma},\mathbf{Z})$, after removing the constant terms, 
problem \eqref{prob:fastbound} can be divided into a set of subproblems on a per-cell basis:
\begin{subequations}\label{prob:upV}
\begin{align}
\underset{\mathbf{V}_{j1},\dots,\mathbf{V}_{jK}}{\text{minimize}}&\quad \sum^K_{k=1}\mathrm{Tr}\Big(2\Re\Big\{\omega_{jk}(\mathbf{I}+\mathbf{\Gamma}_{jk})\mathbf{V}_{jk}^H\mathbf{C}_{jk}^H\mathbf{Y}_{jk}\notag\\
&\quad+\mathbf{V}_{jk}^H(\alpha_{j}\mathbf{I}-\bm\Xi_j)\mathbf{Z}_{jk}\Big\}-\alpha_{j}\mathbf{V}_{jk}^H\mathbf{V}_{jk}\Big)\\
\text{subject to}& \quad \sum^K_{k=1}\mathrm{Tr}(\mathbf{V}_{jk}\mathbf{V}_{jk}^H)\leq P,
\end{align}
\end{subequations}
for $j=1,2,\ldots,L$. The optimal solution of $\mathbf V_{jk}$ is given by
\begin{equation}
    \mathbf{V}_{jk}^\star=\frac{1}{\alpha_j+\eta_j^\star}\bigg(\omega_{jk}\mathbf{C}_{jk}^H\mathbf{Y}_{jk}(\mathbf{I}+\mathbf{\Gamma}_{jk})+\big(\alpha_j\mathbf{I}-\mathbf{\Xi}_j\big)\mathbf{Z}_{jk}\bigg),
\end{equation}
where the optimal Lagrange multiplier $\eta_j^\star$ is still computed as in \eqref{Lagrange multiplier}.
After a bit algebra, it can be shown that the above solution boils down to
\begin{align}\label{eq:upPfast}
\mathbf{V}^\star_{jk}=\left\{
\begin{array}{ll}
    {\bm\Psi}_{jk} & \text{if}\; \sum^K_{k=1}\mathrm{Tr}(\bm\Psi_{jk}^H\bm\Psi_{jk})\leq P\vspace{0.5em}\\
    \beta_j\bm\Psi_{jk} & \text{otherwise},
    \end{array}
    \right.
\end{align}
where
\begin{align}\label{eq:upVnonhomo}
\bm\Psi_{jk}
&=\mathbf{Z}_{jk}+\frac{1}{\alpha_{j}}\Big(\omega_{jk}\mathbf{C}_{jk}^H\mathbf{Y}_{jk}(\mathbf{I}+\mathbf{\Gamma}_{jk})-\mathbf{\Xi}_j\mathbf{Z}_{jk}\Big)
\end{align}
and
\begin{equation}
\beta_j = \sqrt{\frac{P}{\sum^K_{k=1}\mathrm{Tr}(\bm\Psi_{jk}^H\bm\Psi_{jk})}}.
\end{equation}
Algorithm \ref{alg2} summarizes the above steps. Its per-iteration complexity equals $\mathcal{O}(L^2K^2M_r^2M_t^2+LKM_r^3)$.

The following proposition shows that Algorithm \ref{alg2} converges to a stationary point solution not only of problem \eqref{prob:upV} but also of problem \eqref{prob:bound}, so the problem approximation based on Lemma \ref{lemma:nonhomo} does not cause performance loss.
\begin{proposition}
Algorithm \ref{alg2} converges to $(\mathbf V_0,\bm\Gamma_0,\mathbf Y_0,\mathbf Z_0)$ which is a stationary point of problem \eqref{prob:upV}. Moreover, $(\mathbf V_0,\bm\Gamma_0,\mathbf Y_0)$ is a stationary point of problem \eqref{prob:bound}.
\end{proposition}
\begin{IEEEproof}
We only sketch the basic ideas which follow the existing work \cite{shen_accqt} closely. First, Algorithm \ref{alg2} can be recognized as a MM method for problem \eqref{prob:upV}; because each step of Algorithm \ref{alg2} boils down to solving a convex subproblem, the convergence to a stationary point can be verified by the MM theory \cite{shen_accqt}. Moreover, using the inequality in Lemma \ref{lemma:nonhomo} to bound $\hat f(\mathbf{V},\mathbf{\Gamma},\mathbf{Y})$ can be thought of as constructing a surrogate function for problem \eqref{prob:bound}, so each stationary point of problem \eqref{prob:upV} can be mapped to a stationary point of problem \eqref{prob:bound} according to the MM theory.
\end{IEEEproof}

\begin{remark}
Furthermore, our algorithms can be extended to the energy efficiency maximization. Specifically, we first decouple the ratio of the energy efficiency by Dinkelbach's transform \cite{dinkelbach1967nonlinear}. With an auxiliary variable iteratively updated, the new problem is the same as the problem in \eqref{prob:bound}, so the proposed Algorithm \ref{alg1} and Algorithm \ref{alg2} can be readily applied.
\end{remark}

\begin{figure}[t]
    \centering
    \includegraphics[width=0.96\linewidth]{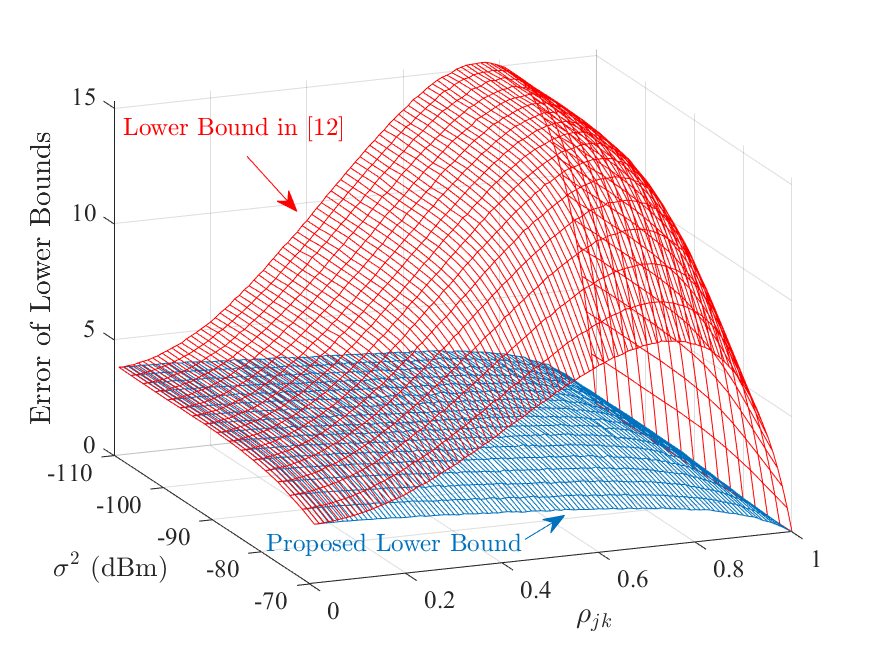}
    \caption{Approximation error between the lower bounds and original objective for different noise power levels and different values of the temporal correlation coefficient $\rho_{jk}$.}
    \label{fig:proposeboundvsrho}
\end{figure}

\section{Simulations}\label{sec:simu}



\subsection{Settings}

We consider two fading models: the Rayleigh fading which is Gaussian, and the Nakagami-$m$ fading which is non-Gaussian. We first describe the Rayleigh fading model from \cite{aal,shi23,swmmse_al}. The channel is modeled as
\begin{align}\label{eq:rayleighmodel}
    \mathbf{H}_{jk,\ell} = \rho_{jk}\bar{\mathbf{H}}_{jk,\ell}+\sqrt{1-(\rho_{jk})^2}\mathbf{W}_{jk,\ell}\odot\mathbf{X}_{jk,\ell},
\end{align}
where $\bar{\mathbf{H}}_{jk,\ell}$ is the static component, $\mathbf{W}_{jk,\ell}\odot\mathbf{X}_{jk,\ell}$ is the random component, and $\rho_{jk}\in(0,1]$ is the tradeoff factor. These parameters are further specified in the following. The real matrix $\mathbf{W}_{jk,\ell}\in\mathbb{R}^{M_r\times M_t}$ reflects on the large-scale fading, each entry of which is computed as $10^{-\xi/10}$ where $\xi=
0.5\times(128.1+37.6\log_{10}d+\tau)$, $d$ denoting the distance in km between the corresponding pair of BS and user, $\tau$ being a zero-mean Gaussian random variable with a standard deviation of 8 dB for the shadowing effect. The fixed matrix $\bar{\mathbf{H}}_{jk,\ell}$ is a realization of $\mathbf{W}_{jk,\ell}\odot\mathbf{X}_{jk,\ell}$ at the beginning of the test.
The random complex matrix $\mathbf{X}_{jk,\ell}\in\mathbb C^{M_r\times M_t}$ has each entry independently drawn from the standard complex Gaussian distribution $\mathcal{CN}(0,1)$. The tradeoff factor $\rho_{jk}$, a.k.a. the temporal correlation coefficient, is generated by Jake's model \cite{jakesmodel}. Specifically, we model it as $\rho_{jk}=J_0(2\pi v_{jk}f_cT/c)$, where $J_0(\cdot)$ is the zeroth order Bessel function of the first kind, $v_{jk}$, $f_c$, $c$, and $T$ are the $(j,k)$-th user's moving speed, the carrier frequency, the speed of light, and the delay after obtaining the static component, respectively. The parameter choice in our simulations follows \emph{3GPP\_3D\_UMa} \cite{3GPP.TR38.901} standard: $f_c=4.8$ GHz, $T=0.5$ ms, $v_{jk}=60$ or $90$ km/h. The corresponding values of $\rho_{jk}$ are 0.8321 or 0.6425 respectively. We consider $1000$ independent blocks and evaluate their average rates.

We also consider the Nakagami-$m$ fading model as in \cite{nakagami}:
\begin{align}\label{eq:nakamodel}
    \mathbf{H}_{jk,\ell} = \rho_{jk}\bar{\mathbf{H}}_{jk,\ell}+\sqrt{1-(\rho_{jk})^2}\mathbf{M}_{jk,\ell}.
\end{align}
Now the random component becomes a Nakagami-$m$ random matrix $\mathbf{M}_{jk,\ell}\in\mathbb C^{M_r\times M_t}$. For the $(m,n)$th entry of $\mathbf{M}_{jk,\ell}$, its phase is uniform on $(0,2\pi]$, and its magnitude is generated independently according to the Nakagami-$m$ distribution $\mathrm{Nakagami}(m,\Omega_{mn})$ \cite{nakagami}, where the shape parameter $m=0.5$ and the mean $\Omega_{mn}$ equals the square of the large-scale fading $\xi$ as defined earlier for the Rayleigh fading model. Again, $\bar{\mathbf{H}}_{jk,\ell}$ is a realization of $\mathbf{M}_{jk,\ell}$ and is fixed ever since. We still average out the performance across $1000$ independent blocks.

We further specify the network setting. The radius of each cell equals $300$ meters. We let $P=30$ dBm and $\omega_{jk}=1$. The noise power is $-90$ dBm by default. For the multi-cell case, we consider 7 wrapped-around cells. We let $M_t=64$, $M_r=2$, and $K=8$ by default for the single-cell case, and $M_t=32$, $K=16$, and $M_r=2$ by default for the multi-cell case. The proposed lower bound in Theorem \ref{thm:lower bound} is compared with the existing lower bound in \cite{gaussianerror1}. Moreover, the benchmark methods for stochastic precoding include:
\begin{itemize}
    \item \emph{WMMSE:} Run the WMMSE algorithm \cite{wmmse_ibc} based on the static components of fading channels.
    \item \emph{Robust Linear Precoding Design (RLPD):} Maximize an approximation of the original objective function \cite{aal}.
    \item \emph{Stochastic WMMSE (SWMMSE):} A model-free precoding method in \cite{swmmse} based on fading channel samples.
\end{itemize}

\begin{figure*}
    \centering
    \subfigure[Single-cell, Rayleigh fading case]{\includegraphics[width=0.48\linewidth]{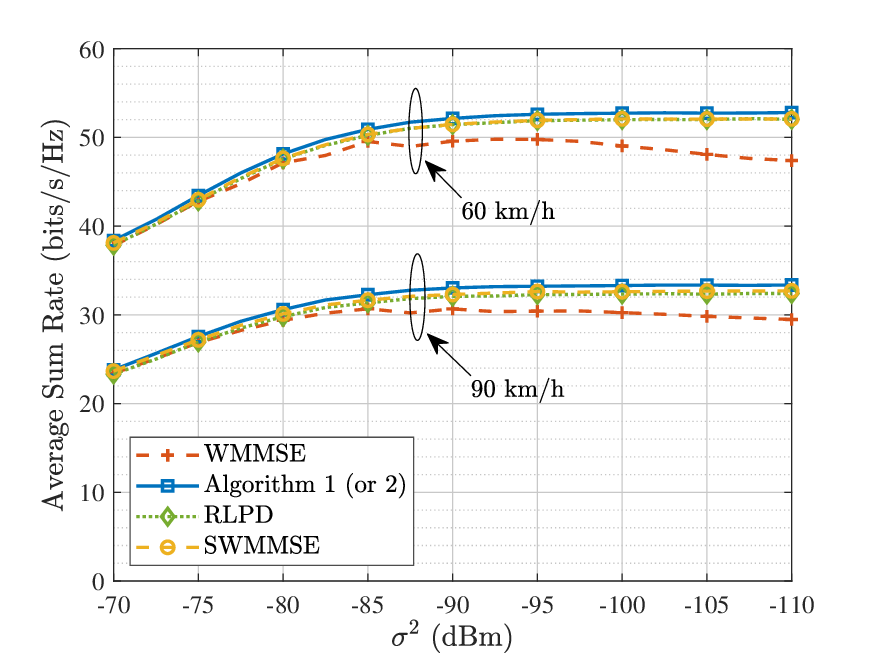}\label{subfig:ervssigma_k8}}
    \subfigure[Multi-cell, Rayleigh fading case]{\includegraphics[width=0.48\linewidth]{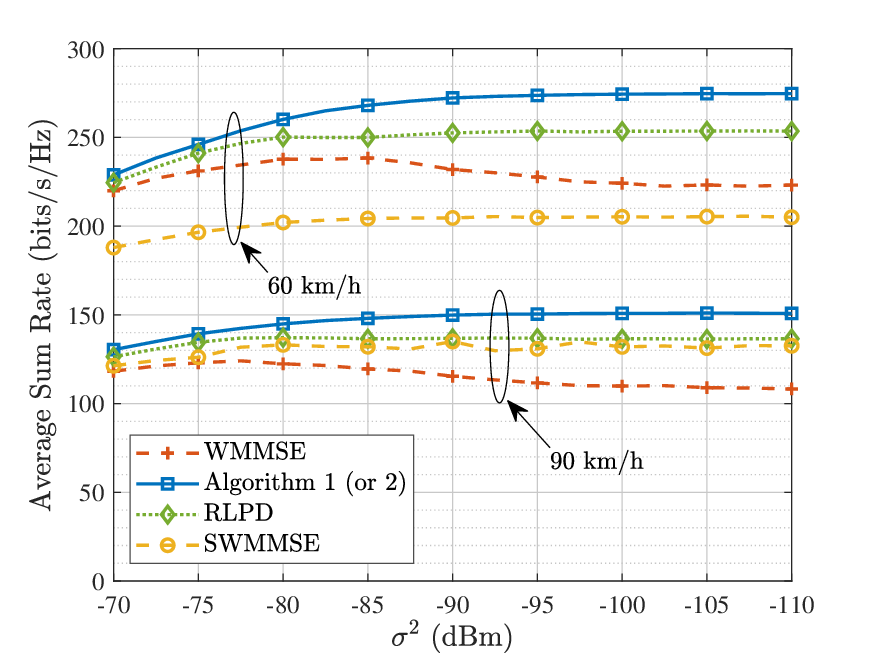}\label{subfig:ervssigma_k16}}
    \subfigure[Single-cell, Nakagami-$m$ fading case]{\includegraphics[width=0.48\linewidth]{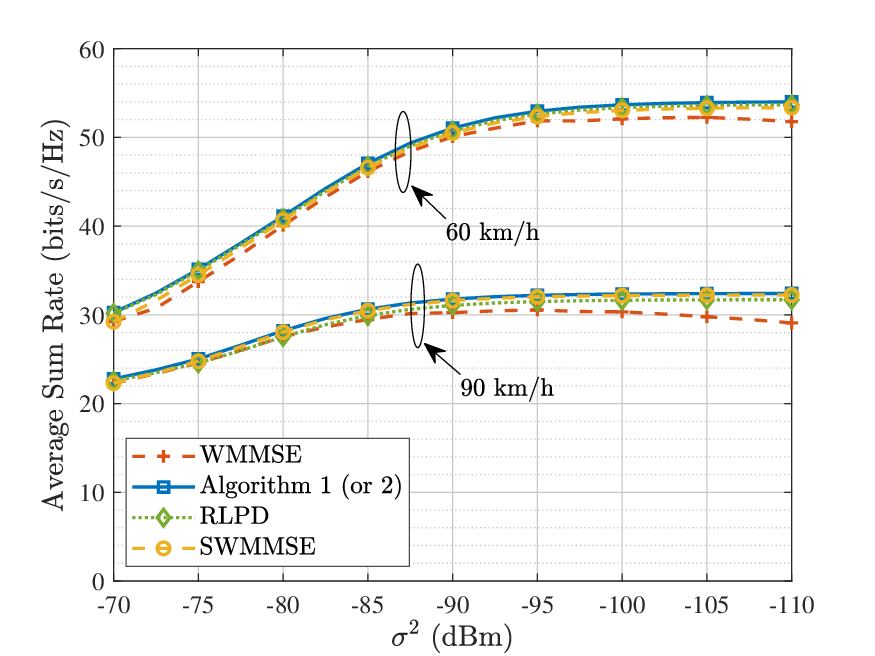}\label{subfig:ervssigma_k8naka}}
    \subfigure[Multi-cell, Nakagami-$m$ fading case]{\includegraphics[width=0.48\linewidth]{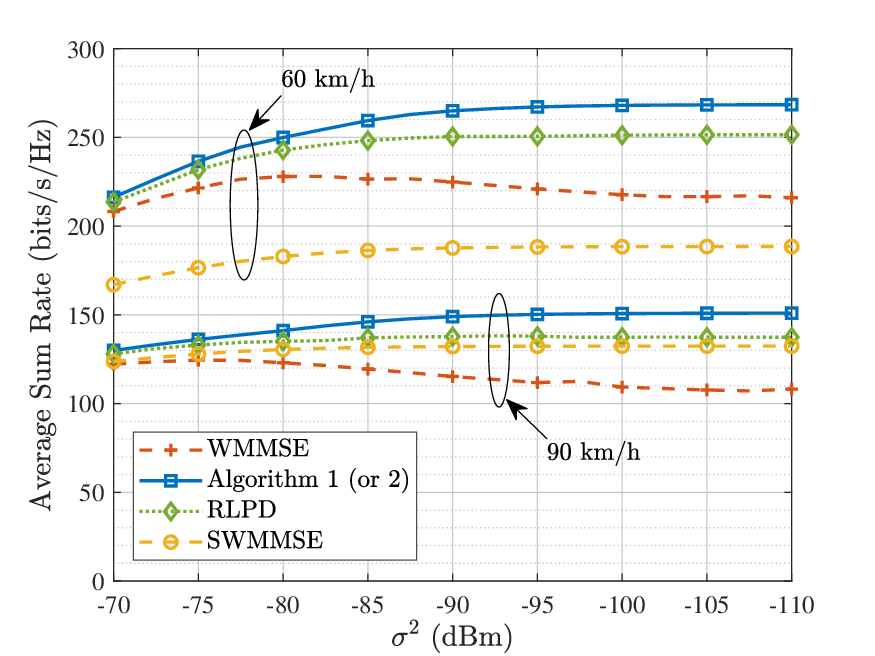}\label{subfig:ervssigma_k16naka}}
    \caption{The long-term average sum rates achieved by the different algorithms under the different fading models and under the different noise power levels when $v=60$ and $90$ km/h.}
    \label{fig:ervssigma}
\end{figure*}
\begin{figure*}
    \centering
    \includegraphics[width=\textwidth]{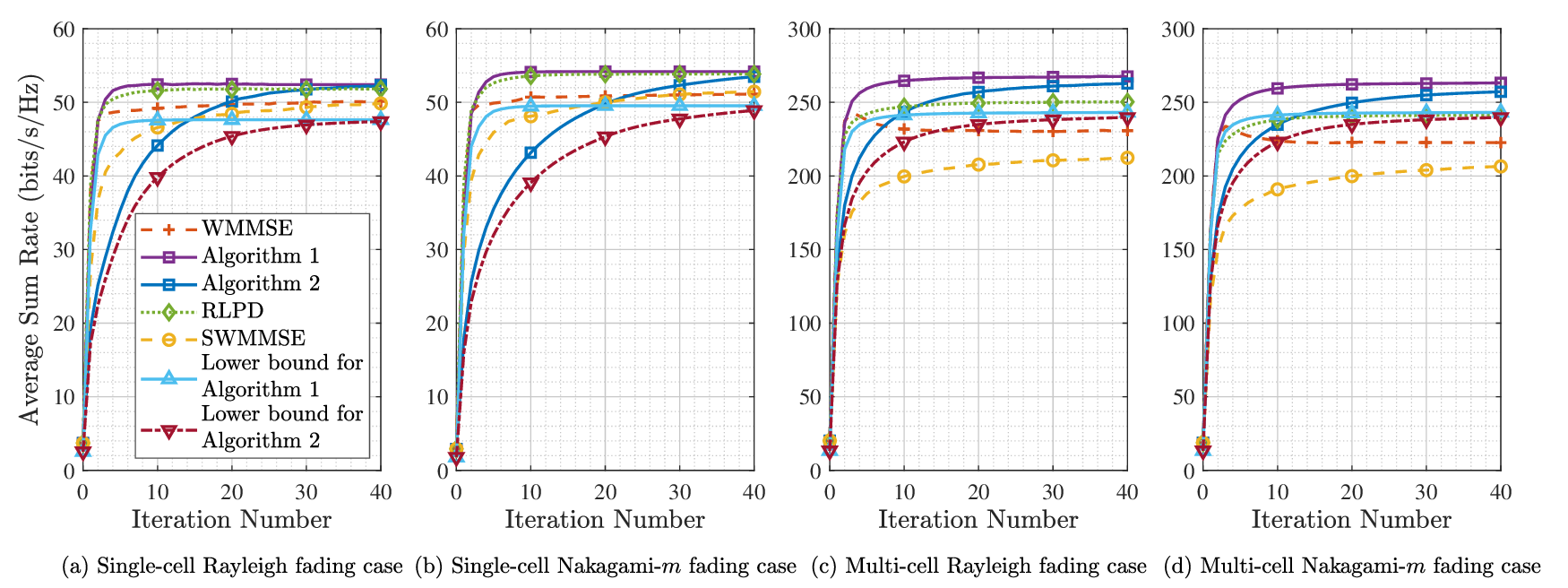}
    \caption{Convergence of different algorithms versus number of iterations under various configurations when $v=60$ km/h.}
    \label{fig:ervsiter_comparison60}
\end{figure*}
\begin{figure*}
    \centering
    \includegraphics[width=\textwidth]{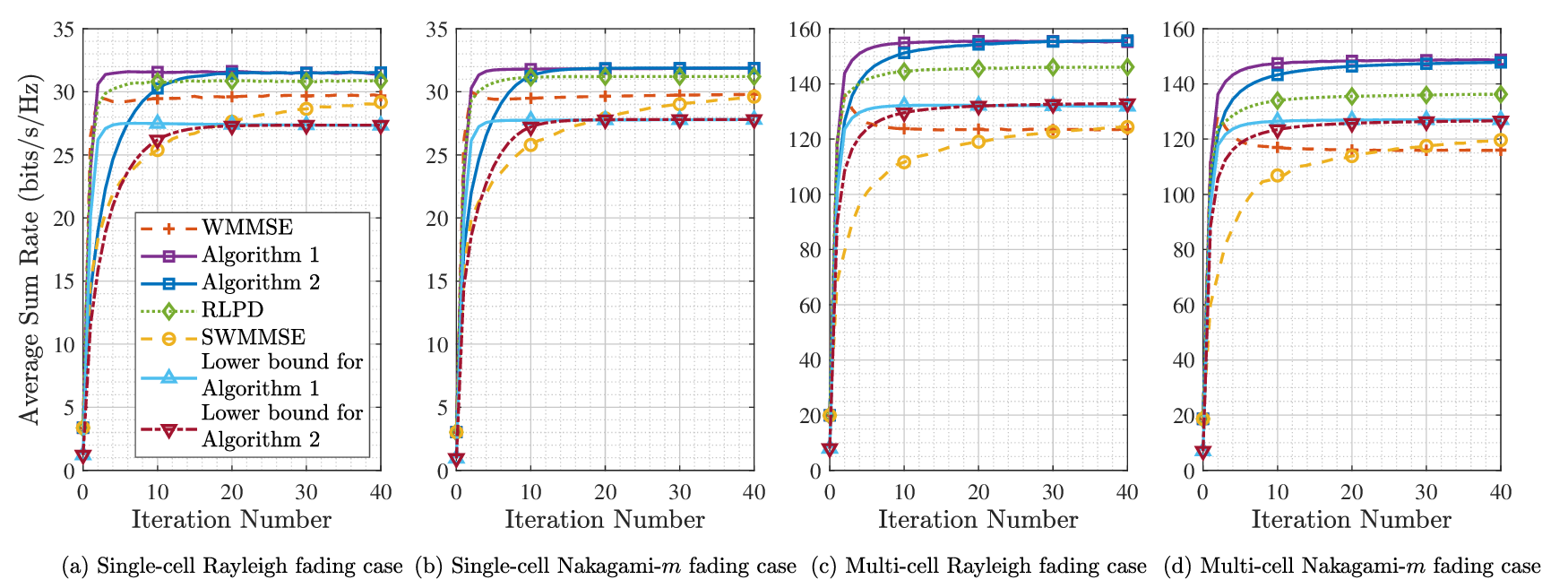}
    \caption{Convergence of different algorithms versus number of iterations under various configurations when $v=90$ km/h.}
    \label{fig:ervsiter_comparison90}
\end{figure*}

\begin{figure*}
    
    \subfigure[Single-cell Rayleigh fading case]{
        \includegraphics[width=0.48\linewidth]{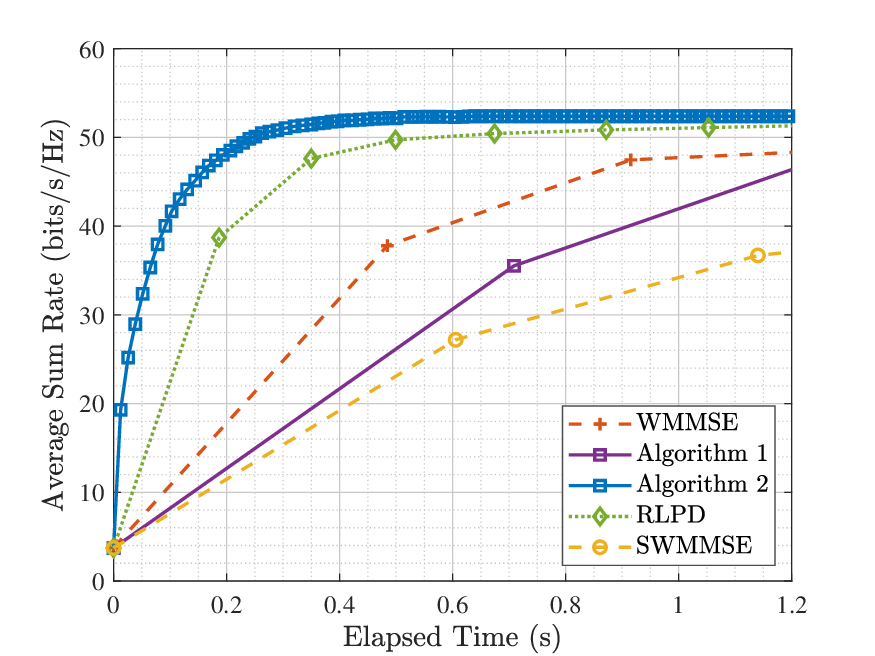}
        \label{subfig:time_sc_k8_v120}}
    \subfigure[Multi-cell Rayleigh fading case]{
        \includegraphics[width=0.48\linewidth]{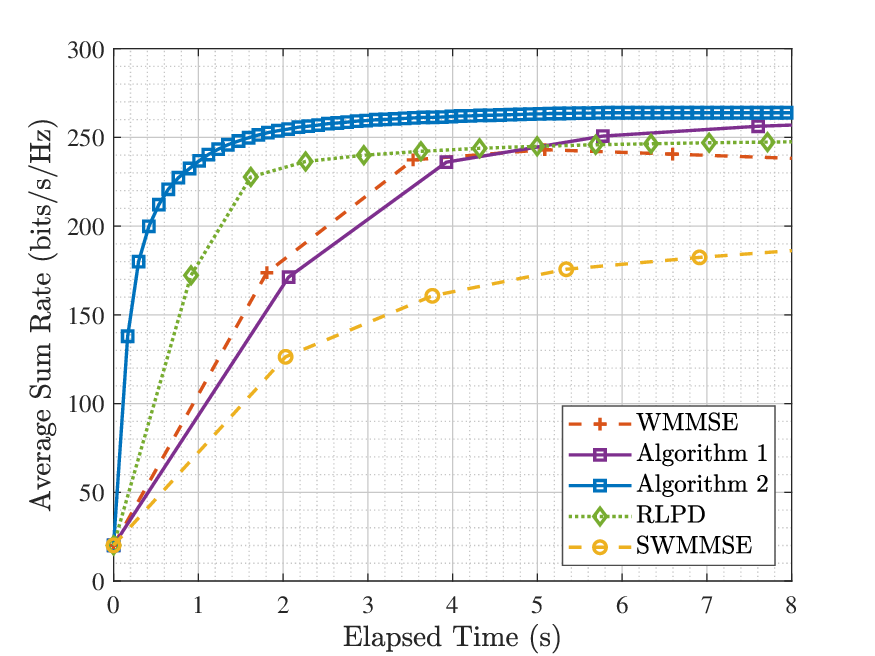}
        \label{subfig:time_mc_k16_v120}}
    \subfigure[Single-cell Nakagami-$m$ fading case]{
        \includegraphics[width=0.48\linewidth]{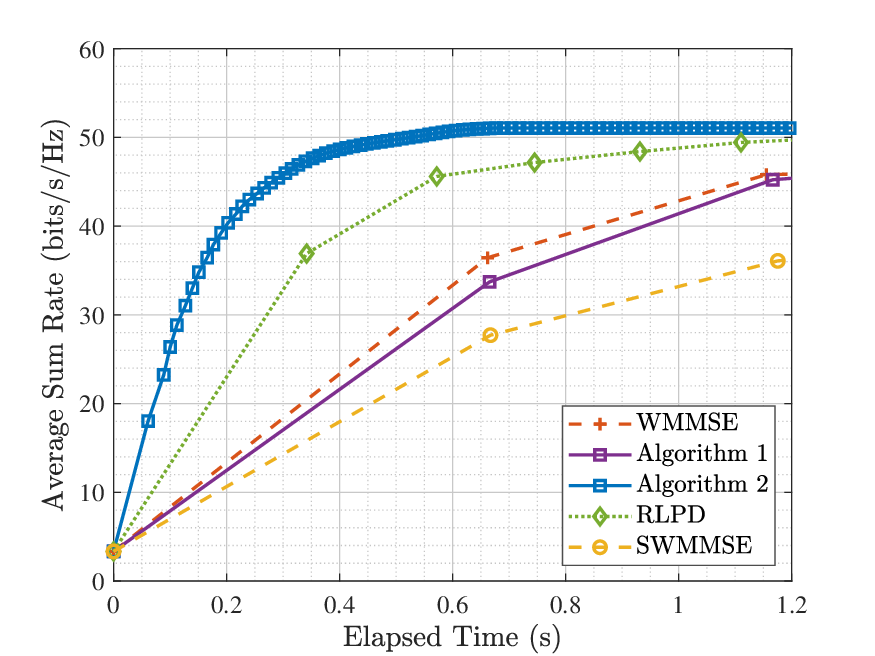}
        \label{subfig:time_sc_k8_v60_naka}}
    \subfigure[Multi-cell Nakagami-$m$ fading case]{
        \includegraphics[width=0.48\linewidth]{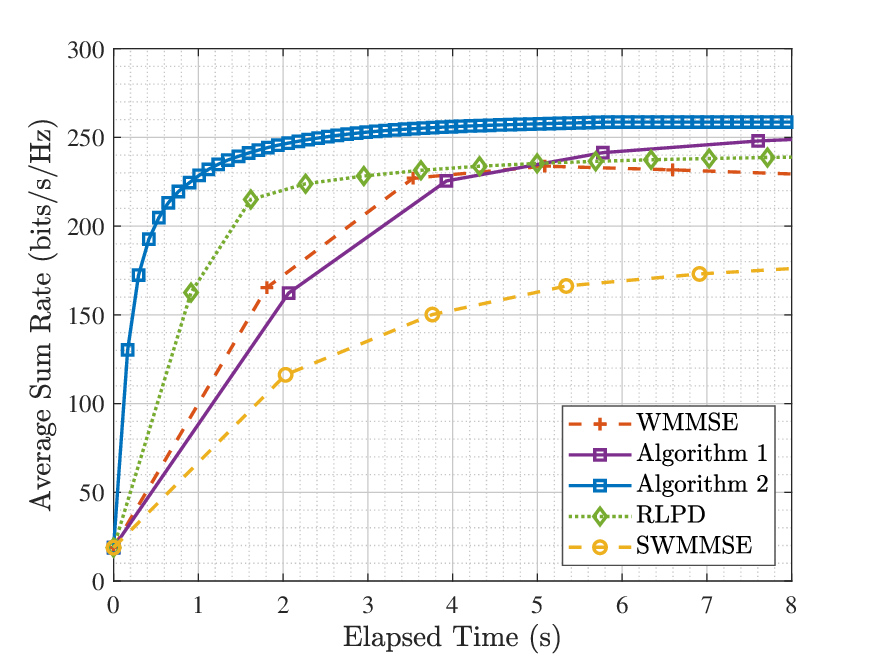}
        \label{subfig:time_mc_k16_v120_naka}}
    
    \caption{Convergence of different algorithms versus elapsed time under various configurations when $v=60$ km/h.}
    \label{fig:ervstime_comparison60}
\end{figure*}
\begin{figure*}
    \centering
    \subfigure[Single-cell Rayleigh fading case]{
        \includegraphics[width=0.48\linewidth]{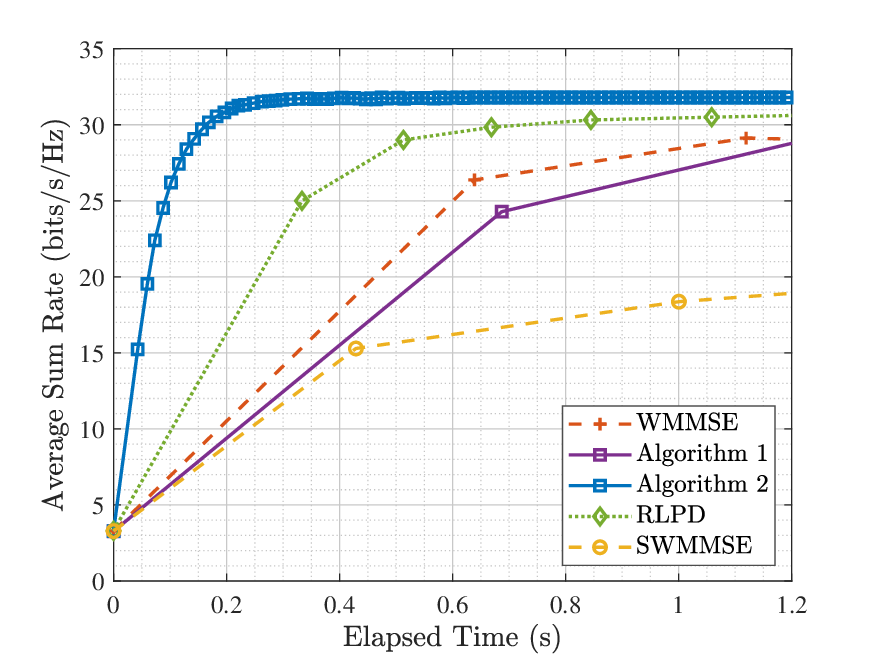}
        \label{subfig:time_sc_k8_v90}}
    \subfigure[Multi-cell Rayleigh fading case]{
        \includegraphics[width=0.48\linewidth]{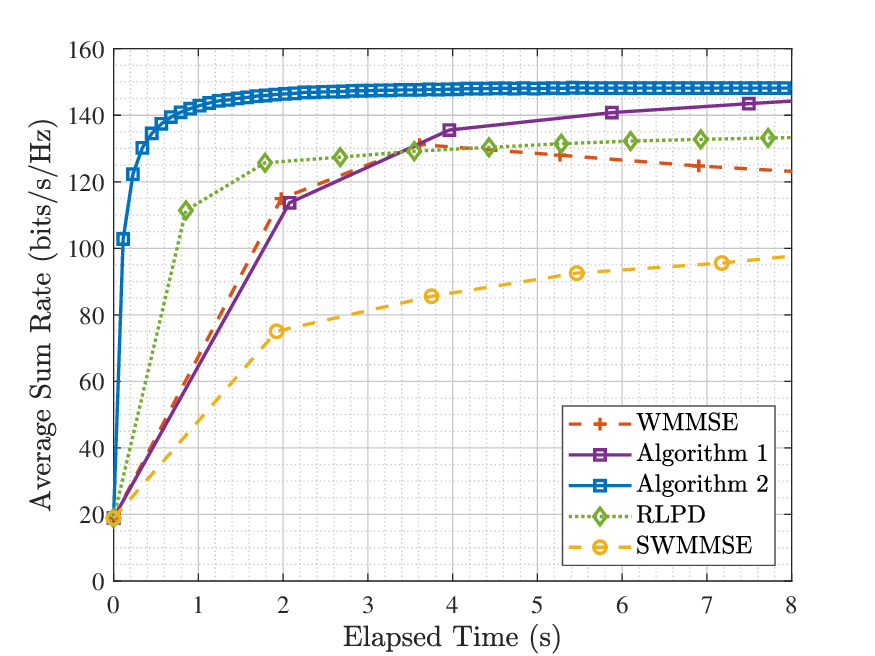}
        \label{subfig:time_mc_k16_v90}}
    \subfigure[Single-cell Nakagami-$m$ fading case]{
        \includegraphics[width=0.48\linewidth]{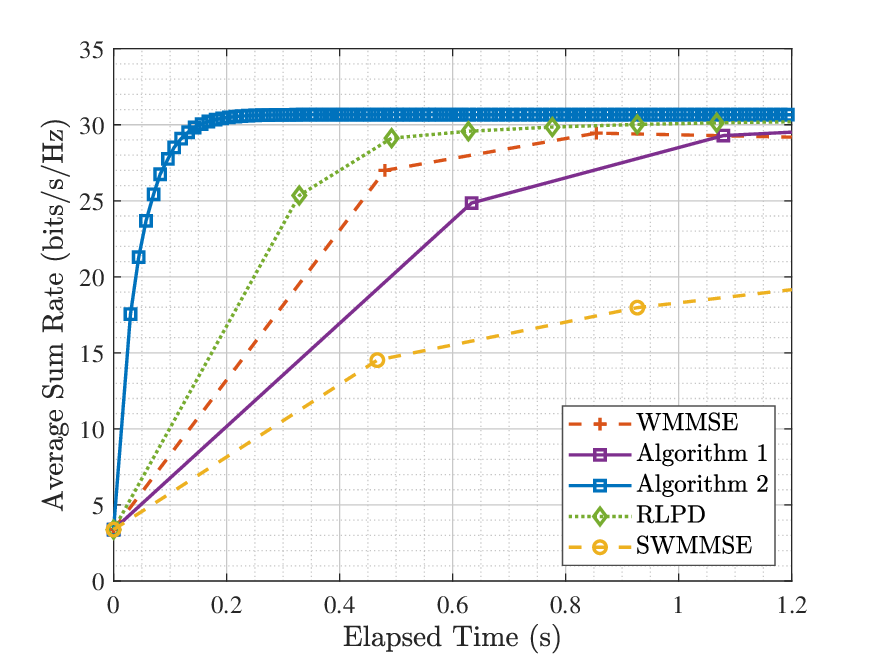}
        \label{subfig:time_sc_k8_v90_naka}}
    \subfigure[Multi-cell Nakagami-$m$ fading case]{
        \includegraphics[width=0.48\linewidth]{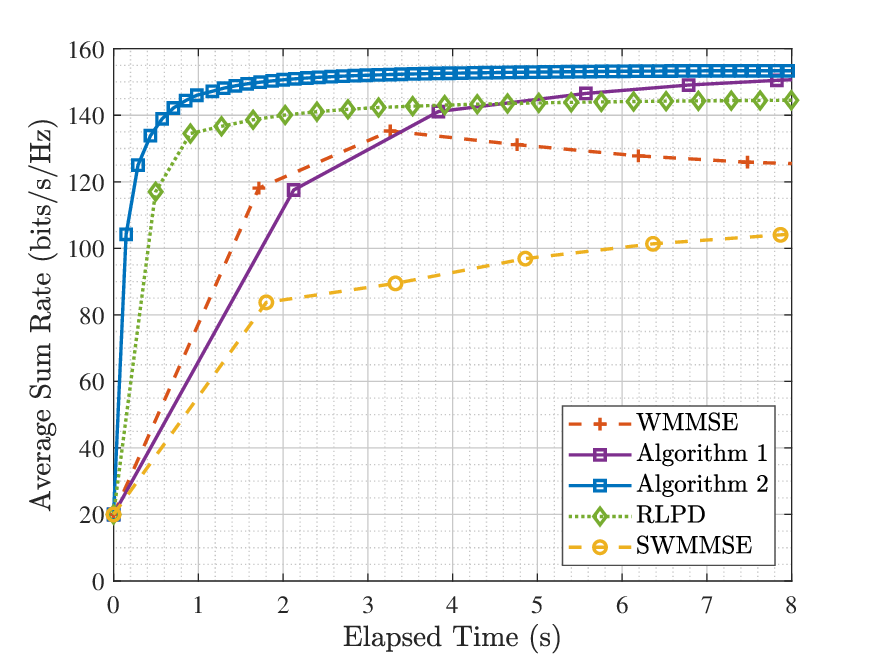}
        \label{subfig:time_mc_k16_v90_naka}}
    \caption{Convergence of different algorithms versus elapsed time under various configurations when $v=90$ km/h.}
    \label{fig:ervstime_comparison90}
\end{figure*}
\begin{figure}
    \centering
    \includegraphics[width=0.96\linewidth]{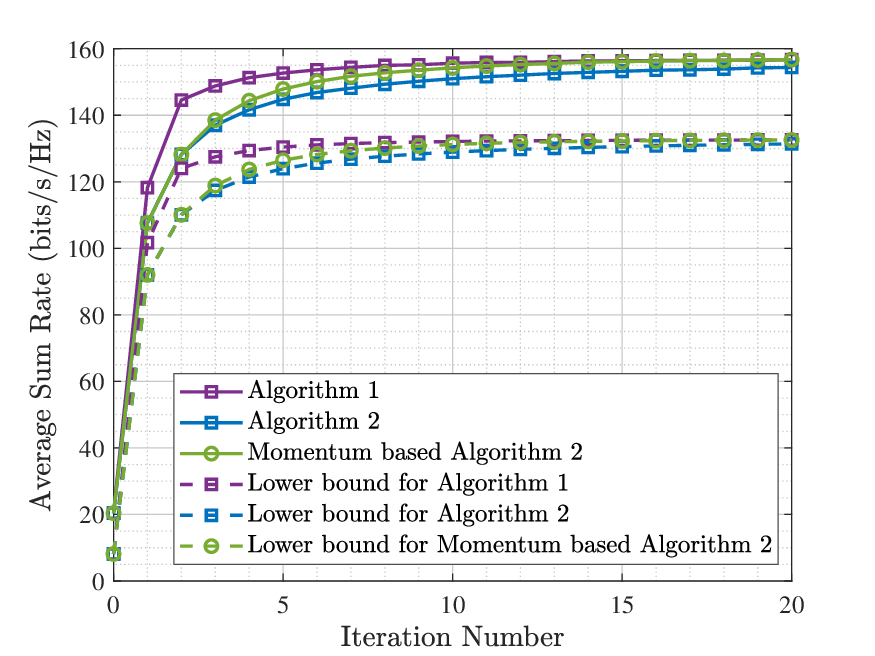}
    \caption{Average sum rates  versus number of iterations under different algorithms or lower bounds in multi-cell system when $v=90$ km/h.}
    \label{fig:momentum}
\end{figure}
\begin{figure}
    \centering
    \includegraphics[width=0.96\linewidth]{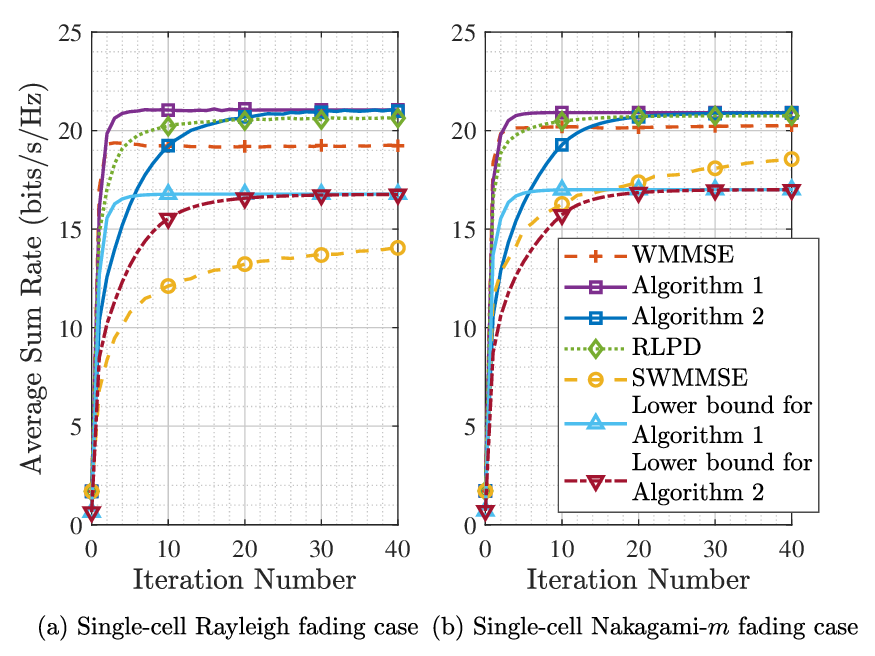}
    \caption{Convergence of different algorithms versus number of iterations under various configurations when $v=900$ km/h and users' weights are not equal.}
    \label{fig:random_weights_iter90}
\end{figure}

\begin{figure*}
    \centering
    \subfigure[Single-cell, Rayleigh fading case]{
        \includegraphics[width=0.48\linewidth]{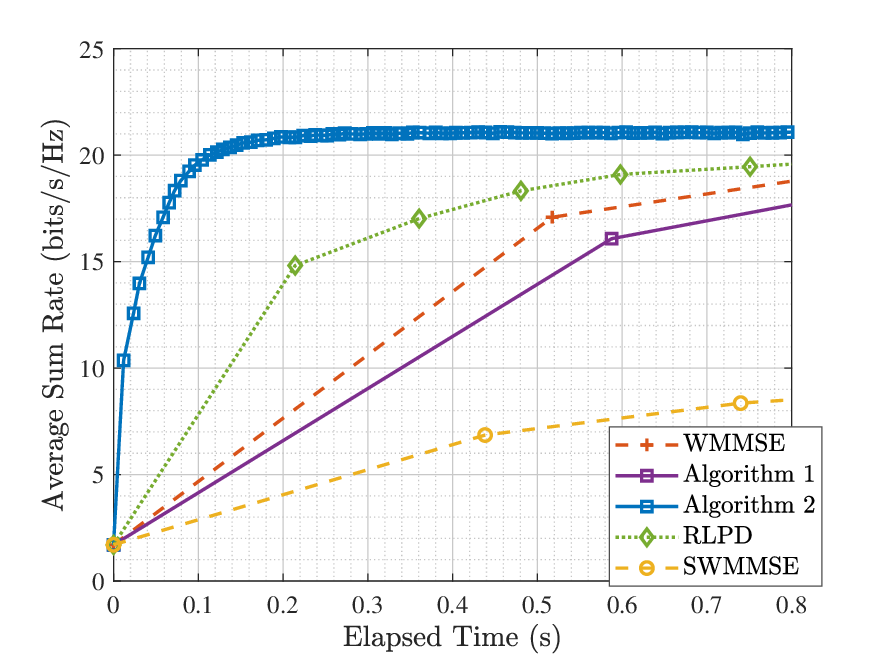}
        \label{subfig:rand_time_sc_k8_v90}}
    \subfigure[Single-cell, Nakagami-$m$ fading case]{
        \includegraphics[width=0.48\linewidth]{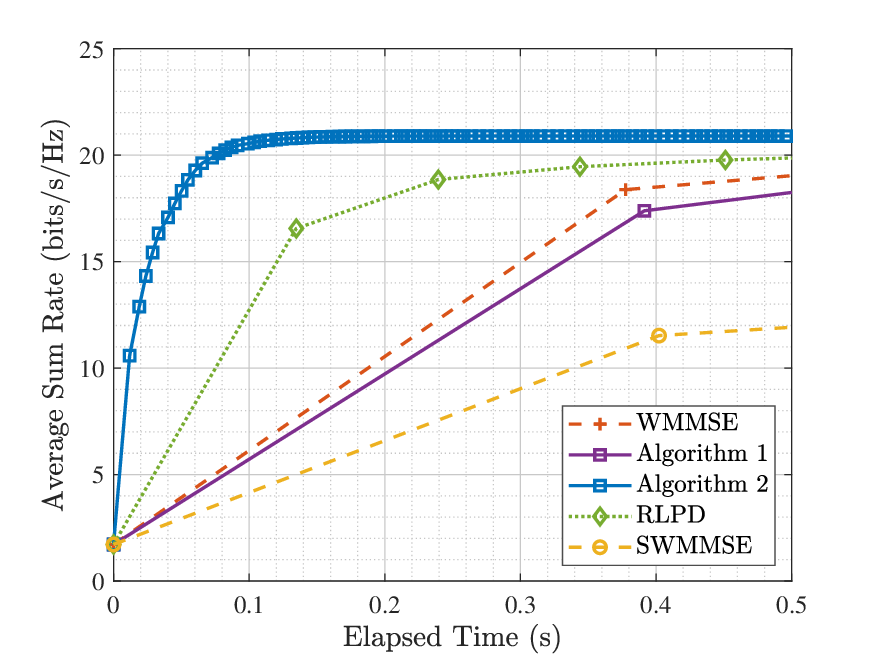}
        \label{subfig:rand_time_sc_k8_v90_naka}}
    \caption{Convergence of different algorithms versus elapsed time under various configurations when $v=90$ km/h and users' weights are not equal.}
    \label{fig:random_weights_time90}
\end{figure*}
\subsection{Test Results}

We begin by comparing different lower bounds in Fig. \ref{fig:proposeboundvsrho}. Consider the Rayleigh fading model. We remark that the proposed lower bound is evaluated with the optimal  $\tilde{\bm\Gamma}^\star$ and $\tilde{\mathbf Y}^\star$, which amounts to the un-parameterized bound $\tilde f(\mathbf V)$ in \eqref{eq:lower_boundftx}. Regarding the deterministic equivalent $\bar f$ in [11], we remark that it is not in closed form. More specifically, $\bar{f}$ depends on a set of auxiliary variables which remain to be updated iteratively. The approximation $\bar f$ equals to the original objective $f$ only after the iterative updates of these auxiliary variables converge. However, it can take infinitely many iterations to attain convergence. We have tried updating these auxiliary variables for 10 iterations (more than [11] did), but the resulting $\bar{f}$ is neither upper bound nor lower bound to $f$. For this reason, we only compare the proposed lower bound with the the existing one in \cite{gaussianerror1}. It can be seen that the proposed bound is much tighter than the existing one in \cite{gaussianerror1}.

Next, we compare the expected sum rates achieved by different algorithms under the different values of noise power $\sigma^2$ in Fig. \ref{fig:ervssigma}. For SWMMSE, we provide a total of $1000$ channel samples for online optimization. Because Algorithm \ref{alg1} and Algorithm \ref{alg2} only differ in the computational complexity, they produce the same performance curves in Fig. \ref{fig:ervssigma}. Observe that the WMMSE algorithm that only depends on the static components of fading channels lead to the worst performance except for the multi-cell scenario with $v = 60 \text{ km/h}$. The proposed method attains much better performance. Its average sum rate is about 5\% higher than that of WMMSE in the single-cell case, and 30\% higher in the multi-cell case. Observe also that SWMMSE yields fairly nice performance in the single-cell case, which is almost equally good as the proposed method. However, SWMMSE becomes much worse in the presence of multiple cells since then it requires many more channel samples. In practice, online optimization by SWMMSE can be quite costly. It is noteworthy that WMMSE outperforms SWMMSE in the multi-cell scenario at $v = 60 \text{ km/h}$. This happens when the following two conditions are both satisfied: (i) the temporal correlation coefficient is large, so that the channels become less random and then the WMMSE is well suited for the case; (ii) there are multiple cells so that the training of the SWMMSE becomes more difficult, and thus the performance of the SWMMSE becomes worse. Thus, the WMMSE rises as the SWMMSE falls. In contrast, for the single-cell scenario, even though the performance of WMMSE still becomes better with a higher temporal correlation coefficient, the stochastic WMMSE can now be well trained and thus can yield good performance. As a result, WMMSE cannot beat the SWMMSE for the single-cell scenario even when the temporal correlation coefficient is large.

Moreover, we compare the speed of convergence for the different algorithms. We now draw Algorithm \ref{alg1} and Algorithm \ref{alg2} separately. First, we evaluate the speed in iterations. As shown in Fig. \ref{fig:ervsiter_comparison60} and Fig. \ref{fig:ervsiter_comparison90}, the convergence curves of the lower bounds are consistent with those of the actual objective function value. Observe also that
WMMSE cannot even yield monotonic improvement for the objective value. It can be seen that SWMMSE converges slowly; this low efficiency can be expected since online optimization data-mines for solution from massive channel samples. In particular, note that Algorithm \ref{alg2} is even slower than Algorithm \ref{alg1}. Actually, this is not a surprising result. When we talk about acceleration in Section \ref{sec:fast FP}, we mean the simplification of each iteration by eliminating matrix inverse. However, as shown in \cite{shen_accqt}, this simplification is at the cost of slowing down the convergence in iterations. Nevertheless, the acceleration in computation outweighs the deceleration in iterations. In plain words, even though it now takes more iterations to converge, the run time of each iteration becomes much shorter, and hence the overall time consumption is still reduced significantly by Algorithm \ref{alg2}. The results from Fig. \ref{fig:ervstime_comparison60} and Fig. \ref{fig:ervstime_comparison90} agree with the above discussion. Observe that Algorithm \ref{alg2} is about 3 times faster than Algorithm \ref{alg1}. Actually, when Algorithm \ref{alg1} just finishes one iteration, Algorithm \ref{alg2} already reaches convergence. We further compare their run times in Table \ref{tab:run_time_comm} by using different numbers of antennas. Observe that the efficiency advantage of Algorithm \ref{alg2} over Algorithm \ref{alg1} becomes larger when the number of antennas increases, so Algorithm \ref{alg2} is well suited for the large-scale MIMO network.

\begin{table}[t]
\footnotesize
    \renewcommand{\arraystretch}{1.3}
\centering
\caption{\small Per-Iteration Run Time of Different Methods.}
\begin{tabular}{lrrrr}
\firsthline
& \multicolumn{4}{c}{Average Running Time (second)}\\
\cline{2-5}
Algorithm      & $M_t=32$ & $M_t=64$ & $M_t=128$ & $M_t=256$\\
\hline
WMMSE          & 0.0960 & 0.2735 & 1.1187 & 5.4836 \\
Algorithm \ref{alg1}  & 0.1209 & 0.3707 & 1.4573 & 6.8598\\
Algorithm \ref{alg2}   & 0.0039  & 0.0064 & 0.0167 & 0.1597\\
RLPD           & 0.0656 & 0.1555 & 0.4953 & 3.0894 \\
SWMMSE         & 0.1199 & 0.3974 & 1.4926 & 5.0732 \\

\lasthline
\end{tabular}
\label{tab:run_time_comm}
\end{table}

Furthermore, following \cite{shen_accqt}, we accelerate Algorithm \ref{alg2} by incorporating \emph{momentum} into it. As shown in Fig. \ref{fig:momentum}, momentum can indeed improve the convergence of  Algorithm \ref{alg2}. Both Algorithm~\ref{alg1} and Algorithm~\ref{alg2} guarantee convergence to a stationary point of the lower bound problem. Since Algorithm~\ref{alg2} is based on a looser bound, its convergence slows down, but its per-iteration complexity is lower because of the cheaper subproblem. Thus, Algorithm.~\ref{alg2} is preferable when the computational power is highly limited and the full convergence is not required.


Lastly, we consider the case in which the rate weight varies from user to user. For simplicity, we focus on the single-cell network with the user moving speed $v=90$ km/h. The rate weights of the eight users are $\{0.617, 0.094, 0.171, 0.887, 0.886, 0.251, 0.102, 0.772\}$. The simulation results are summarized in Fig. \ref{fig:random_weights_iter90} and Fig. \ref{fig:random_weights_time90}. It can be seen that the distinct-weight case does not differ much from the unit-weight case as formerly shown in Fig. \ref{fig:ervsiter_comparison90} and Fig. \ref{fig:ervstime_comparison90}. Thus, the proposed algorithm accounts for various rate weight settings.

\section{Conclusion}\label{sec:conclu}
This work investigates the stochastic precoding design by using the first and second moments of fading channels. As compared to the existing works that assume a particular channel distribution, our method has the advantage of working for a generic fading model; as compared to the existing works that adopt a model-free approach, our method has the advantage of providing more efficient optimization. The heart of our method lies in a stochastic extension of the existing matrix FP methods. Although this paper focuses on the downlink network, the proposed algorithm can be readily adapted to other network topologies such as device-to-device, uplink, and cell-free.

\bibliographystyle{IEEEtran}
\bibliography{IEEEabrv,strings}                       
\begin{IEEEbiographynophoto}{Wenyu Wang}(Graduate Student Member, IEEE) received the B.E. degree in digital media technology from Huazhong University of Science and Technology in 2020. He is currently working toward the Ph.D degree with the School of Science and Engineering, The Chinese University of Hong Kong (Shenzhen), China. His research interests include intelligent reflecting surface, ultra-reliable and low-latency communication and fractional programming.
\end{IEEEbiographynophoto}
\begin{IEEEbiographynophoto}{Kaiming Shen} (Senior Member, IEEE) received the B.Eng. degree in information security and the B.Sc. degree in mathematics from Shanghai Jiao Tong University, China in 2011, and then the Ph.D. degree in electrical and computer engineering from the University of Toronto, Canada in 2020. He has been with the School of Science and Engineering at The Chinese University of Hong Kong (Shenzhen), China as a tenure-track assistant professor since 2020. His research interests include optimization, wireless communications, and information theory. He currently serves as an Editor for IEEE Transactions on Wireless Communications. He is a member of the Signal Processing for Communications and Networking (SPCOM) Technical Committee of the IEEE Signal Processing Society. He received the IEEE Signal Processing Society Young Author Best Paper Award in 2021, the University Teaching Achievement Award in 2023, the Frontiers of Science Award in 2024, and the Chinese Information Theory Society Young Researcher Award in 2025.
\end{IEEEbiographynophoto}
\end{document}